\documentclass[11pt,letterpaper]{article}
\usepackage{graphicx}
\usepackage[margin=2.57cm]{geometry}
\usepackage{subcaption}
\usepackage{amssymb}
\usepackage{amsmath}
\usepackage{amsfonts}
\usepackage{theorem}
\usepackage{natbib}
\usepackage{hyperref}
\usepackage{algorithm}
\usepackage{algorithmic}
\usepackage{times}
\usepackage{epstopdf}
\usepackage{filemod}
\usepackage{etoolbox}
\usepackage{varioref}

\restylefloat{table}
\hypersetup{backref,dvips,bookmarks=true,pdfauthor=Ju-Chen Justin Yang,colorlinks=false,breaklinks=true,hyperfigures=true,pdfstartview=FitH,linkcolor=blue,citecolor=black}

\theorembodyfont{\upshape}

\AtEndEnvironment{remark}{$\Diamond$}

\def\bi{\begin{itemize}}
\def\ei{\end{itemize}}

\allowdisplaybreaks
\graphicspath{{../../graphs/}}
\input{tcilatex}
\begin{document}

\title{Fast and optimal nonparametric sequential design for astronomical
observations}
\author{ \textsc{Justin J. Yang} \\
\textit{Department of Statistics,}\\
\textit{Harvard University}\\
\texttt{juchenjustinyang@fas.harvard.edu} \and \textsc{Xufei Wang} \\
\textit{Department of Statistics,}\\
\textit{Harvard University}\\
\texttt{xufeiwang@fas.harvard.edu} \and \textsc{Pavlos Protopapas} \\
\textit{Institute for Applied Computational Science,}\\
\textit{Harvard School of Engineering and Applied Sciences}\\
\texttt{pavlos@seas.harvard.edu} \and \textsc{Luke Bornn} \\
\textit{Department of Statistics,}\\
\textit{Harvard University}\\
\texttt{bornn@stat.harvard.edu}}
\maketitle

\begin{abstract}
The spectral energy distribution (SED) is a relatively easy way for
astronomers to distinguish between different astronomical objects such as
galaxies, black holes, and stellar objects. By comparing the observations
from a source at different frequencies with template models, astronomers are
able to infer the type of this observed object. In this paper, we take a
Bayesian model averaging perspective to learn astronomical objects,
employing a Bayesian nonparametric approach to accommodate the deviation
from convex combinations of known log-SEDs. To effectively use telescope
time for observations, we then study Bayesian nonparametric sequential
experimental design without conjugacy, in which we use sequential Monte
Carlo as an efficient tool to maximize the volume of information stored in
the posterior distribution of the parameters of interest. A new technique
for performing inferences in log-Gaussian Cox processes called the Poisson
log-normal approximation is also proposed. Simulations show the speed,
accuracy, and usefulness of our method. While the strategy we propose in
this paper is brand new in the astronomy literature, the inferential
techniques developed apply to more general nonparametric sequential
experimental design problems.
\end{abstract}



\baselineskip=20pt

\section{Introduction}

Spectral energy distributions (SEDs), as well as their fitting, are used in
many branches of astronomy to characterize astronomical sources. For
example, photometric redshift estimation (distance estimation of sources)
relies heavily on the SED morphology of galaxies. Because of its strong
predictive power and relative ease of use, SED fitting is very commonly used
in astronomy.

Our \emph{scientific goal} in this paper is to fit telescope observations to
various existing template SEDs, which are generated either from observations
of known astronomical objects or models, in order to (i) classify a new
astronomical source, (ii) analyze a new blended source as a geometrically
weighted average of template models, or (iii) detect the evidence of a new
type of SED which cannot be directly described by known templates.

However, due to the high cost of using sophisticated telescopes and the
limitation of observation time, it is necessary to carefully design the
observational strategy using all the information we have, including the
template models, the specifications of telescope filters\footnote{%
Filters here mean the physical filters used with the detectors on the
telescope in order to restrict the observed electromagnetic bandwidth, a
range of frequencies.}, and those existing observations. In the context of
SED fitting, this is equivalent to specify the set of filters to use in
order to better achieve the aforementioned scientific goal. Decisions
regarding these specifications must be made before the data collection.
Because specific information is usually available prior to the use of the
telescope, the Bayesian framework will play an important role.

Especially, sequential design is preferred as opposed to a non-sequential
one. The following three advantages motivate our choice of this methodology:
Firstly, the optimal sequential design procedure must be at least as good as
a fixed design procedure (\cite{Chaloner&Verdinelli-1995}). Secondly, it is
usually more computationally efficient, as finding the optimal
non-sequential design for all design variables at a time is usually NP-hard (%
\cite{Ko&Lee&Queyranne-1995}). Finally, a sequential design can also
incorporate the existing literatures on multi-armed bandit problems (\cite%
{Robbins-1952,Berry&Fristedt-1985,Krause&Ong-2011}) and sequential Monte
Carlo (SMC) (\cite{Cherkassky&Bornn-2013}).

Our \emph{statistical goal} in this paper is to provide an efficient and
fast calculation scheme to reach the optimal sequential design under the SED
fitting context. One computational difficulty comes from the incorporation
of a \emph{non-conjugate} Bayesian nonparametric prior on the deviation
between the (convex combinations of) template models and the truth. It is
this non-conjugate setting that makes our work different from the existing
literatures in \emph{Bayesian nonparametric sequential experimental design
(BNS-ED)}, which usually assumes a Gaussian conjugacy.

There are several major contributions we make in this paper. We are the
first to our knowledge to study BNS-ED without conjugacy. Our second
contribution is to provide a fast SMC algorithm to solve the general BNS-ED
problems. Furthermore, by employing the special model structure of
log-Gaussian Cox process (LGCP), the main model of interest in this paper,
we introduce a new technique called Poisson log-normal approximation (PoLNA)
to improve computation speed. As a third contribution, we apply the new
methodology stated above to sequentially choose the best filters to use and
fit the SED on-the-fly. To show how our method could be applied, we perform
simulation tests on real astronomical templates and demonstrate that, using
our algorithm, one can better analyze the unknown SED in terms of template
models with fewer observations, which shows the practical value of our
methodology.

The rest of this paper is organized as follows. In Section \ref{motivation},
we formulate the main scientific problem of interest and specify the
quantities and notations we will use throughout this paper. In Section \ref%
{SMC}, we introduce the statistical model, design perspective, and general
SMC inferential scheme for BNS-ED. We then introduce the specialized
technique PoLNA for LGCP in Section \ref{PoLNA}. Simulation examples are
provided in Section \ref{Simu.}. Finally, we discuss several related works
and conclude in Section \ref{Conc.}. We leave more detailed calculations to
the Supplementary Material.

\section{Motivation\label{motivation}}

\begin{figure}[t]
\centering
\begin{subfigure}[b]{0.48\textwidth}
    \centering\includegraphics[width=\linewidth]{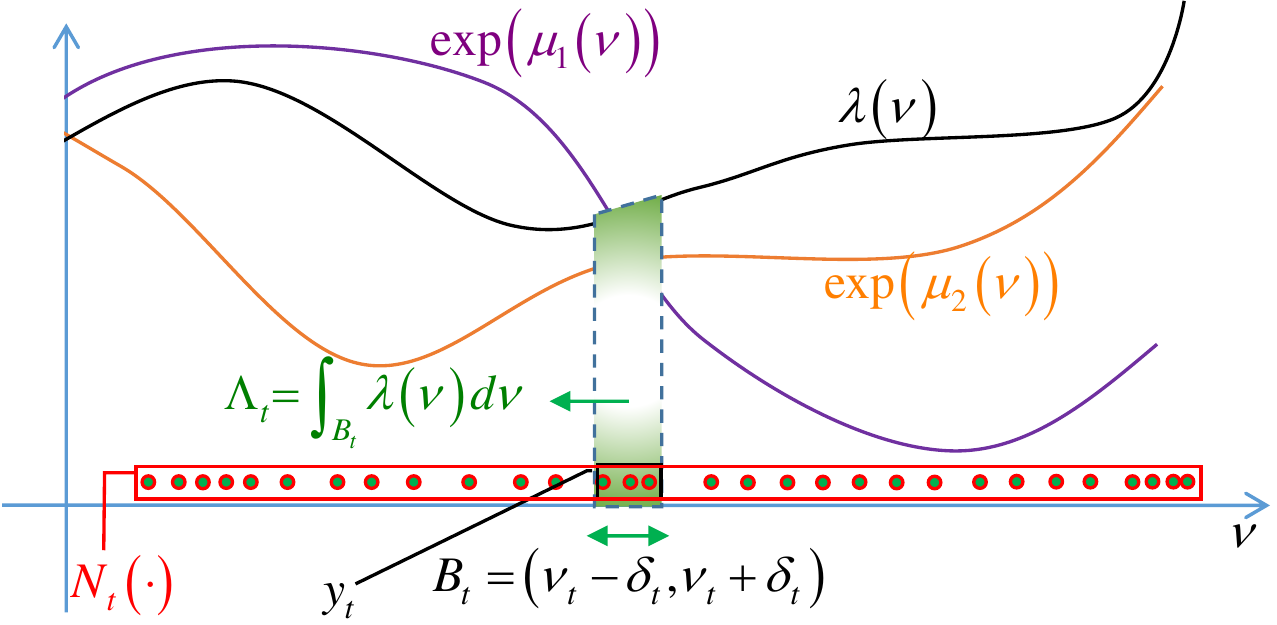}
    \caption{
        Relationships between different quantities involved.
        Blue and orange lines mean some other SED templates.
    }\label{sed_fitting_graph}
\end{subfigure}
\begin{subfigure}[b]{0.48\textwidth}
    \centering\includegraphics[width=\linewidth]{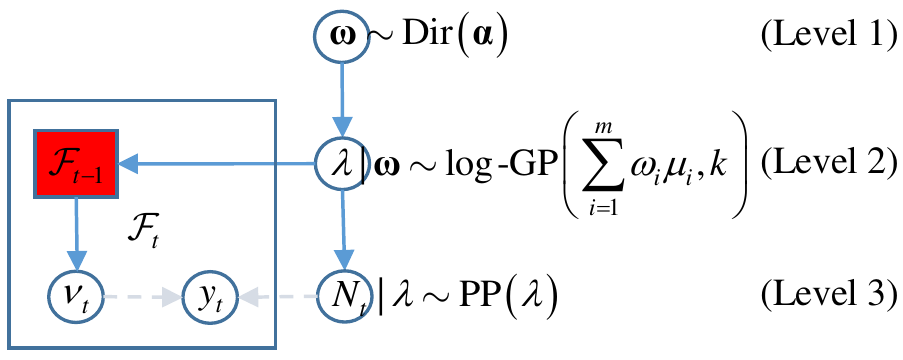}
    \caption{
        A three-level hierarchical (mixture) log-Gaussian Cox process.
        The dashed array here means a deterministic relationship.
    }\label{sed_model}
\end{subfigure}
\caption{ Graphical illustrations for the problem of SED fitting as well as
its model setting. }
\end{figure}

A graphical illustration for the problem of fitting a spectral energy
distribution (SED) is shown in Figure \ref{sed_fitting_graph}. 

The true but unknown SED from an astronomical source is denoted as $\lambda
\left( \nu \right) $, where $\nu $ represents the frequency of the photons
and $\lambda \left( \nu \right) $ means the intensity of those arriving
photons with frequency $\nu $. Several known templates are denoted as $\exp
\left( \mu _{i}\left( \nu \right) \right) $ , where $\mu _{i}$'s are the
template log-SEDs. The actual observation $y_{t}$ we collect from the
telescope is the total number of photons observed using a particular filter
with a certain bandwidth $B_{t}=\left( \nu _{t}-\delta _{t},\nu _{t}+\delta
_{t}\right) $, which centers around $\nu _{t}$ with a frequency range $%
\delta _{t}$. Let $N_{t}\left( \cdot \right) $ denote the complete empirical
distribution of photon arrivals emitted at time $t$ with different
frequencies. Then $y_{t}=N_{t}\left( B_{t}\right) $, because one can only
observe the total counts of photons in the bandwidth $B_{t}$.

\textbf{Motivating astronomical problem.} Use all of the collected data $%
y_{t}$'s as well as the filters $B_{t}$'s to describe the unknown SED $%
\lambda \left( \nu \right) $ in terms of the existing SED templates, $\exp
\left( \mu _{i}\left( \nu \right) \right) $'s.

Most existing astronomy literatures use frequentist model selection
methodologies to determine which template could best represent the truth,
that is, to find a unique $i$ such that $\lambda \left( \nu \right) =\exp
\left( \mu _{i}\left( \nu \right) \right) $. However these methods (i) fail
to quantify the uncertainty of this sole template selection away from the
true SED, (ii) cannot take advantage of such uncertainty information to
suggest the next step of observational setting, and (iii) ignore the
possibility that the true SED might not be accounted for by those selected
templates, as shown by the region of larger $\nu $ in Figure \ref%
{sed_fitting_graph}.

To address (ii), a naive and prevailing approach for collecting observations
is to use each single available filter on the telescope and fit the SED
until the completion of data collection. However, this might be expensive
because of the time constraint on telescope availability (on average only 8
night hours per day). Multiple visits to the telescope due to an inefficient
observational strategy will increase the monetary cost for astronomers.

A unifying and adequate solution addressing all of the issues above is
obtained through a Bayesian approach. Instead of choosing a sole template,
we take a Bayesian model averaging perspective on those templates by
assuming that $\lambda \left( \nu \right) =\exp \left( \sum_{i=1}^{m}\omega
_{i}\mu _{i}\left( \nu \right) \right) $, where $\mathbf{\omega }\triangleq
\left( \omega _{1},...,\omega _{m}\right) ^{T}$ is the vector of mixture
weights assigned to each template such that $\sum_{i=1}^{m}\omega _{i}=1$.
We summarize the existing observations as a prior distribution on $\mathbf{%
\omega }$ and accordingly induce the posterior distribution on $\mathbf{%
\omega }$ by incorporating our current observations. For (i), the
uncertainty information for the goodness-of-fit of the data could be
extracted from the posterior distribution on $\mathbf{\omega }$, e.g., the
posterior probability intervals for each $\omega _{i}$. For (ii), the
posterior distribution on $\mathbf{\omega }$ also serves as the primary
proxy for choosing the next filter to use. This is how Bayesian sequential
experimental design comes into play.

Finally, to address (iii), we can write $\lambda \left( \nu \right) =\exp
\left( \sum_{i=1}^{m}\omega _{i}\mu _{i}\left( \nu \right) +\epsilon \left(
\nu \right) \right) $, where $\epsilon \left( \nu \right) $ describes the
deviation between the true log-SED and (convex combinations of) those
selected templates. Due to the limited knowledge of the unstructured term $%
\epsilon \left( \nu \right) $, we naturally impose a weak prior on it, which
requires the use of a Bayesian nonparametric prior and hence motivates our
study of Bayesian nonparametric sequential experimental design (BNS-ED).

\section{Design and Inferential Scheme for BNS-ED\label{SMC}}

In this Section, we will specify the statistical models as well as the
utility function for BNS-ED. We then discuss the inference for BNS-ED in
general without employing any specific features of our model. The
derivations of equations (\ref{Bayesian_Updating}), (\ref{EIG_2}), and (\ref%
{EIG_3}) can be found in Supplementary Materials \ref{Derivations}.

\subsection{Model Specification}

Due to the discrete nature of our observations, it is convenient to model
the photon arrivals at different frequencies, $N_{t}$, as an inhomogeneous
Poisson point process (or Poisson random measure):%
\begin{eqnarray*}
N_{t}\left( \cdot \right) &\backsim &_{|\lambda }\mathrm{PP}\left( \lambda
\right) \text{ i.i.d. for all }t\text{,} \\
y_{t} &\triangleq &N_{t}\left( B_{t}\right) \backsim _{|B_{t},\lambda }%
\mathrm{Poisson}\left( \int_{B_{t}}\lambda \left( \nu \right) \mathrm{d}\nu
\right) .
\end{eqnarray*}%
Also, without making any strong assumptions, it is convenient to assign the
prior distribution for $\epsilon \left( \nu \right) $---in a nonparametric
way---as a Gaussian process (GP)\footnote{%
The choice of GP here is purely conventional. It is possible to replace this
nonparametric prior by any stable process (e.g. a Cauchy process) if
alternative tail behavior is desired. Our SMC approach will easily scale to
such alternative specifications.}: $\epsilon \left( \nu \right) \backsim 
\mathrm{GP}\left( 0,k\right) $, where $k=k\left( \nu ,\nu ^{\prime }\right) $
is the covariance function (or kernel function). In this paper, we assume
that $k$ is given, fixed, and coming from a parametric family. We can also
conduct a full Bayesian inference on the parameters of $k$ by using the
inferential techniques we introduce in this paper, but for the ease of
demonstration we choose not to address this point further.

Later on in the simulation we will conventionally make a stronger assumption
that $k\left( \nu ,\nu \right) =\sigma ^{2}$ for any $\nu $, which might not
be true in practice. To have a more accurate representation of the reality
about $k$, we can perform a maximum likelihood fitting for $k$ from a
flexible covariance structure based on existing observations. One anonymous
reader also mentioned that $\sigma ^{2}$ should be small for observing a
well-understood object, as the deviation term is essentially unnecessary.
Hence, even though our Bayesian framework will allow the flexibility that
the truth can deviate from the selected templates, it will also be
compatible with the frequentist methodologies used by astronomers, which
could provide strong predictive power for the astronomical objects.

We finally assign a prior on $\mathbf{\omega }$ as $\mathbf{\omega }\backsim 
\mathrm{Dir}\left( \mathbf{\alpha }\right) $ by incorporating some astronomy
prior knowledge, which we choose not to discuss here for the clarity of
presentation. Another way of choosing $\mathbf{\alpha }$ is from an
empirical moment matching on existing observations. In practice, $m$ might
be very large, so we could also choose $\mathbf{\alpha }$ to encourage
sparse $\mathbf{\omega }$.

Thus, in terms of a Bayesian graphical (or hierarchical) model
representation, shown in Figure \ref{sed_model}, the main parameter of
interest---the unknown SED $\lambda $---has a nonparametric prior
distribution that is a log-Gaussian process with a mean function depending
on $\mathbf{\omega }$ and a known covariance function. Levels 1--3 together
specify a mixture (over $\mathbf{\omega }$) log-Gaussian Cox process (LGCP)
on $N_{t}\left( \cdot \right) $ (see \cite%
{Ghahramani&Griffiths-2006,Lawrence&Moore-2007,Rue&Martino-2009} for its
generalization), which we will use throughout this paper as the main model
of interest.

\subsection{Design Objective}

Following the seminal work by \cite{Lindley-1956}, we consider the expected
gain in Shannon information (\cite{Shannon-1948}) as the utility function.
Precisely, our design goal in BNS-ED is to sequentially choose a design that
maximizes the expected Kullback-Leibler divergence between the posterior
distributions of $\mathbf{\omega }$ at time $t$ and $t-1$: 
\begin{equation}
EIG_{t}\left( B_{t}\right) \triangleq \mathbb{E}_{y_{t}|B_{t},\mathcal{F}%
_{t-1}}\left( D_{\mathrm{KL}}\left( p\left( \mathbf{\omega }|y_{t},B_{t},%
\mathcal{F}_{t-1}\right) ||p\left( \mathbf{\omega }|\mathcal{F}_{t-1}\right)
\right) \right) ,  \label{EIG}
\end{equation}%
where $\mathcal{F}_{t-1}\triangleq \sigma \left(
y_{1},B_{1},...,y_{t-1},B_{t-1}\right) $ denotes all the historical
information before the $t$-th observation. To summarize, we want to have a
series of design decisions that extract the most information for $\mathbf{%
\omega }$ after each observation.

Note that here we do not study the information gain directly on $\lambda $
since our astronomical goal in this paper is more to analyze the unknown
truth with the existing templates. We wish that the inclusion of the
deviation term in some cases (like Example 2 in Section \ref{Simu.}) can
provide some signals of the existence of deviation, but, when there is no
deviation, we still wish to get more information about $\mathbf{\omega }$.

By using 
\begin{equation}
p\left( \mathbf{\omega }|\mathcal{F}_{t}\right) =\dfrac{p\left( y_{t}|B_{t},%
\mathcal{F}_{t-1},\mathbf{\omega }\right) }{p\left( y_{t}|B_{t},\mathcal{F}%
_{t-1}\right) }p\left( \mathbf{\omega }|\mathcal{F}_{t-1}\right) ,
\label{Bayesian_Updating}
\end{equation}%
the expected information gain in equation (\ref{EIG}) can be further
rewritten as 
\begin{equation}
EIG_{t}\left( B_{t}\right) =\mathbb{E}_{\mathbf{\omega }|\mathcal{F}%
_{t-1}}\left( D_{\mathrm{KL}}\left( p\left( y_{t}|B_{t},\mathcal{F}_{t-1},%
\mathbf{\omega }\right) ||p\left( y_{t}|B_{t},\mathcal{F}_{t-1}\right)
\right) \right) .  \label{EIG_2}
\end{equation}%
The problem now reduces to approximating the expectation in equation (\ref%
{EIG_2}), which will be approached with sequential Monte Carlo techniques.

\subsection{Sequential Monte Carlo (SMC) Inference}

We employ the SMC procedure to approximate each of the posterior
distributions $p\left( \mathbf{\omega }|\mathcal{F}_{t-1}\right) $ by a set
of particles $\left\{ \mathbf{\omega }_{t-1}^{\left( i\right) },\psi
_{t-1}^{\left( i\right) }\right\} _{i=1}^{N}$. Thus, to approximate the
expected information gain in equation (\ref{EIG_2}), we can use 
\begin{equation}
EIG_{t}\left( B_{t}\right) \approx \sum_{i=1}^{N}\psi _{t-1}^{\left(
i\right) }\sum_{y_{t}=0}^{\infty }\log \left( \dfrac{p\left( y_{t}|B_{t},%
\mathcal{F}_{t-1},\mathbf{\omega }_{t-1}^{\left( i\right) }\right) }{%
\sum_{i^{\prime }=1}^{N}\psi _{t-1}^{\left( i^{\prime }\right) }p\left(
y_{t}|B_{t},\mathcal{F}_{t-1},\mathbf{\omega }_{t-1}^{\left( i^{\prime
}\right) }\right) }\right) p\left( y_{t}|B_{t},\mathcal{F}_{t-1},\mathbf{%
\omega }_{t-1}^{\left( i\right) }\right) ,  \label{EIG_3}
\end{equation}%
where the summation over $y_{t}$ can be further narrowed as we describe in
Subsection \ref{Sum_range_of_y}.

Based on this approximation, we choose $B_{t}$ to maximize the approximated
expected information gain from the set of available filters ($B_{t}=\limfunc{%
argmax}\limits_{B\in \mathrm{Filters}}EIG_{t}\left( B\right) $). Using this
filter, we acquire a new observation $y_{t}$. Then, according to equation (%
\ref{Bayesian_Updating}), we update the particles at time $t$ via 
\begin{equation*}
\mathbf{\omega }_{t}^{\left( i\right) }=\mathbf{\omega }_{t-1}^{\left(
i\right) },\ \psi _{t}^{\left( i\right) }\propto \psi _{t-1}^{\left(
i\right) }\times p\left( y_{t}\left\vert B_{t},\mathcal{F}_{t-1},\mathbf{%
\omega }_{t-1}^{\left( i\right) }\right. \right) ,
\end{equation*}%
for $i=1,...,N$. Every time the particles are updated as above, we monitor
the effective sample size of the particles. Once the effective sample size
drops below a given threshold, we resample from the current particles.
Furthermore, in order to increase the diversity of the particles, we perturb
each resampled particle by a Markovian move. To achieve this, we first note
that 
\begin{equation*}
p\left( \mathbf{\omega }|\mathcal{F}_{t}\right) \propto
\prod_{s=1}^{t}p\left( y_{s}|B_{s},\mathcal{F}_{s-1},\mathbf{\omega }\right)
p\left( \mathbf{\omega }\right) .
\end{equation*}%
Then, to sample from $p\left( \mathbf{\omega }|\mathcal{F}_{t}\right) $, we
just propose a new $\mathbf{\omega }_{t}^{\left( i\right) ,\ast }$ for each $%
i=1,...,N$ by drawing from $\mathrm{Dir}\left( \tau \mathbf{\omega }%
_{t}^{\left( i\right) }\right) $, where $\tau $ is a tuning parameter
representing the step size, and then accept the proposal with the usual
Metropolis-Hastings acceptance probability. Algorithm \ref{Main-Alg} in the
Supplementary Materials \ref{Sec.:SMC alg.} describes the complete
methodology.

\section{Efficient Computations for LGCP\label{PoLNA}}

The SMC procedure described above relies heavily on a fast and accurate way
to calculate the following posterior predictive distribution for $y_{t}$
using the filter $B_{t}$: 
\begin{equation}
p\left( y_{t}|B_{t},\mathcal{F}_{t-1},\mathbf{\omega }\right) =\dfrac{\int
p\left( y_{t}|B_{t},\eta \right) \prod_{s=1}^{t-1}p\left( y_{s}|B_{s},\eta
\right) p\left( \eta |\mathbf{\omega }\right) \mathrm{d}\eta }{%
\prod_{s=1}^{t-1}p\left( y_{s}|B_{s},\mathcal{F}_{s-1},\mathbf{\omega }%
\right) },  \label{marginal_like}
\end{equation}%
where $\eta \triangleq \log \left( \lambda \right) $. This might be achieved
by using a vanilla Monte Carlo estimate (see \cite{Meeds&Welling-2014} for a
recent development of this kind of simulation-based technique). However, due
to the functional nature of a Gaussian process, the sample space in
calculating equation (\ref{marginal_like}) is too large for vanilla Monte
Carlo integration to be computationally efficient. Furthermore, this large
scale Monte Carlo estimate needs to be repeated $N$ times for each
observation time $t$, which clearly slows down the performance of Algorithm %
\ref{Main-Alg}. An alternative calculation of the posterior predictive
distribution will be introduced in the following two Subsections. We call
this new approach the \emph{Poisson log-normal approximation (PoLNA)} (see
e.g. \cite{Adams&Murrary&MacKay-2009} and \cite%
{Simpson&Illian&Lindgren&Sorbye&Rue-2013} for some common ways to infer
LGCP). In the third Subsection, we apply PoLNA to reduce the amount of
computations needed to calculate the Kullback-Leibler divergence.

\subsection{Poisson Log-Normal Approximation (PoLNA)}

The key idea of this approximation is to reduce the dimensionality of the
integral in equation (\ref{marginal_like}) by finding the joint distribution
of 
\begin{equation*}
\mathbf{\Lambda }=\left( \Lambda _{1},...,\Lambda _{t}\right) ^{T}\triangleq
\left( \Lambda \left( B_{1},\eta \right) ,...,\Lambda \left( B_{t},\eta
\right) \right) ^{T}\triangleq \left( \int_{B_{1}}e^{\eta \left( \nu \right)
}\mathrm{d}\nu ,...,\int_{B_{t}}e^{\eta \left( \nu \right) }\mathrm{d}\nu
\right) ^{T},
\end{equation*}%
so we will have 
\begin{equation*}
\int p\left( y_{t}|B_{t},\eta \right) \prod_{s=1}^{t-1}p\left(
y_{s}|B_{s},\eta \right) p\left( \eta |\mathbf{\omega }\right) \mathrm{d}%
\eta =\int \mathrm{Poisson}\left( y_{t}\left\vert \Lambda _{t}\right.
\right) \prod_{s=1}^{t-1}\mathrm{Poisson}\left( y_{s}\left\vert \Lambda
_{s}\right. \right) p\left( \mathbf{\Lambda }|\mathbf{\omega }\right) 
\mathrm{d}\mathbf{\Lambda }.
\end{equation*}%
Simulation studies (where a simple example for $t=1$ is shown in Figure \ref%
{Fig.: PoLNA fundamental} of Supplementary Materials \ref{get_par_multi})
indicate that the joint distribution of $\log \mathbf{\Lambda }$ can be
approximated by a multivariate normal distribution $\mathcal{N}_{t}\left( 
\mathbf{\mu }_{\Lambda },\mathbf{\Sigma }_{\Lambda }\right) $ with a high
accuracy\footnote{%
For notation clarity, in this Subsection we allow $\mathbf{\Sigma }_{\Lambda
}$ to be degenerate, i.e., some of the $\Lambda _{s}$'s might have
correlation $1$.} , where $\mathbf{\mu }_{\Lambda }$ and $\mathbf{\Sigma }%
_{\Lambda }$ can be obtained either from a Monte Carlo estimate or from a
deterministic numerical calculation using the extended technique of \cite%
{Safak-1993}. The detailed calculation for the later proposal can be found
in the Supplementary Materials \ref{get_par_multi}.

Hence, by plugging into this approximation, the posterior predictive
distribution becomes the conditional distribution of a multivariate Poisson
log-normal distribution, $\mathrm{PLN}\left( \mathbf{\mu },\mathbf{\Sigma }%
\right) $, which has a joint probability mass function%
\begin{equation}
\mathrm{PLN}\left( \mathbf{y}|\mathbf{\mu },\mathbf{\Sigma }\right) =\int
\prod_{s=1}^{t}\mathrm{Poisson}\left( y_{s}\left\vert \Lambda _{s}\right.
\right) \mathcal{N}_{t}\left( \log \mathbf{\Lambda }|\mathbf{\mu },\mathbf{%
\Sigma }\right) \mathrm{d}\left( \log \mathbf{\Lambda }\right) .
\label{PLN_PMF}
\end{equation}%
The (multivariate) Poisson log-normal distribution has several tractable
properties such as analytical formulas for its mean vector or covariance
matrix, unimodal feature, and subexponentially decaying tail. It has been
studied in depth by \cite{Aitchison&Ho-1989} and \cite{Perline-1998}.

\subsection{Laplace Transform Approximation of Multivariate Log-Normal
Distribution}

Equation (\ref{PLN_PMF}) is a low dimensional integral (usually there are
only $10$ different filters available on the telescope), so one common way
to achieve this type of numerical integration is from the \emph{multivariate
Gaussian Hermite quadrature}, which does not take full advantage of the
special form of the integrand---a product of several Poisson likelihoods.
Thus, in this Subsection, we will introduce an alternative way to
approximate equation (\ref{PLN_PMF}) using an approximation to the Laplace
transform of a multivariate log-normal distribution.

First, we rewrite (\ref{PLN_PMF}) as 
\begin{equation}
\mathrm{PLN}\left( \mathbf{y}|\mathbf{\tilde{\mu}},\mathbf{\tilde{\Sigma}}%
\right) =\dfrac{e^{\mathbf{S}^{T}\mathbf{\tilde{\Sigma}S}/2+\mathbf{\tilde{%
\mu}}^{T}\mathbf{S}}}{\Pi _{s=1}^{n}\Pi _{n=1}^{n_{s}}\tilde{y}_{sn}!}\int
e^{-\mathbf{n}^{T}\mathbf{\tilde{\Lambda}}}\log \mathcal{N}_{u}\left( \left. 
\mathbf{\tilde{\Lambda}}\right\vert \mathbf{\tilde{\mu}}+\mathbf{\tilde{%
\Sigma}S},\mathbf{\tilde{\Sigma}}\right) \mathrm{d}\mathbf{\tilde{\Lambda}},
\label{PLN_as_Laplace_Trans}
\end{equation}%
where $\tilde{\Lambda}_{s}$'s are the unique components of $\mathbf{\Lambda }
$ with corresponding multivariate log-normal parameters, $\mathbf{\tilde{\mu}%
}$ and $\mathbf{\tilde{\Sigma}}$, and 
\begin{eqnarray*}
\tilde{y}_{sn} &\backsim &_{|\tilde{\Lambda}_{s}}\mathrm{Poisson}\left( 
\tilde{\Lambda}_{s}\right) \ \text{i.i.d. for all }n=1,...,n_{s},\ s=1,...,u;
\\
\mathbf{y} &=&\left. \left[ \tilde{y}_{sn}\right] _{n=1}^{n_{s}}\right.
_{s=1}^{u};\ \mathbf{S}\triangleq \left( \sum_{n=1}^{n_{1}}\tilde{y}%
_{1n},...,\sum_{n=1}^{n_{u}}\tilde{y}_{un}\right) ^{T};\ \mathbf{n}%
\triangleq \left( n_{1},...,n_{u}\right) ^{T}.
\end{eqnarray*}%
Thus, the multivariate Poisson log-normal distribution is fully
characterized by the Laplace transform of a multivariate log-normal
distribution.

Employing the same technique presented by \cite%
{Asmussen&Jensen&Rojas-Nandayapa-2013} (we omit the detailed proof here), we
can derive a sharp approximation to the Laplace transform of a multivariate
log-normal distribution as 
\begin{equation}
\int e^{-\mathbf{n}^{T}\mathbf{\tilde{\Lambda}}}\log \mathcal{N}_{u}\left(
\left. \mathbf{\tilde{\Lambda}}\right\vert \mathbf{\tilde{\mu}}+\mathbf{%
\tilde{\Sigma}S},\mathbf{\tilde{\Sigma}}\right) \mathrm{d}\mathbf{\tilde{%
\Lambda}}\approx \dfrac{\exp \left( -\dfrac{1}{2}\left( 
\begin{array}{c}
\mathbf{W}_{u}\left( \mathbf{M}\right) ^{T}\mathbf{\tilde{\Sigma}}^{-1}%
\mathbf{W}_{u}\left( \mathbf{M}\right) \\ 
+\mathbf{1}_{u}^{T}\mathbf{\tilde{\Sigma}}^{-1}\mathbf{W}_{u}\left( \mathbf{M%
}\right)%
\end{array}%
\right) \right) }{\sqrt{\det \left( \mathbf{I}_{u}+\mathbf{M}\func{diag}%
\left( e^{\mathbf{-W}_{u}\left( \mathbf{M}\right) }\right) \right) }},
\label{Laplace_of_LN}
\end{equation}%
where $\mathbf{M}\triangleq \mathbf{\tilde{\Sigma}}\func{diag}\left( \mathbf{%
n}\right) \func{diag}\left( e^{\mathbf{\tilde{\mu}}+\mathbf{\tilde{\Sigma}S}%
}\right) $ and $\mathbf{W}_{u}\left( \mathbf{M}\right) $ is the multivariate
Lambert W function defined as the unique solution of $\mathbf{M}\exp \left( -%
\mathbf{W}_{u}\left( \mathbf{M}\right) \right) =\mathbf{W}_{u}\left( \mathbf{%
M}\right) $. This approximation is derived via the Laplace approximation in
an asymptotic sense but it stays sharp over the entire domain of convergence
of the Laplace transform. Combining equations (\ref{PLN_as_Laplace_Trans})
and (\ref{Laplace_of_LN}) gives us an accurate and fast approximation to the
multivariate Poisson log-normal distribution.

\subsection{Efficient Calculation of Kullback-Leibler Divergence\label%
{Sum_range_of_y}}

In this Subsection, we will propose an efficient calculation scheme for
equation (\ref{EIG_3}) by limiting the Kullback-Leibler divergence
calculations over an effective range of $y_{t}$.

For a given particle $\mathbf{\omega }_{t-1}^{\left( i\right) }$ and a
filter $B_{t}$, the inner summation in equation (\ref{EIG_3}) can be well
approximated by summing only over those $y_{t}$ in the $\left( 1-\alpha
\right) $-interval of $p\left( y_{t}\left\vert B_{t},\mathcal{F}_{t-1},%
\mathbf{\omega }_{t-1}^{\left( i\right) }\right. \right) $---thanks to the
unimodal property and the subexponentially decaying tail of a Poisson
log-normal distribution. We usually set $\alpha =5\%$. The remaining
question now is how to find these two quantiles for the conditional Poisson
log-normal distribution. We will derive a pair of conservative bounds by
focusing on the univariate Poisson log-normal distribution $\mathrm{PLN}%
\left( y_{t}|\mu _{tt},\Sigma _{tt}\right) $ since the unconditional
distribution will be fatter than the conditional one.

As \cite{Perline-1998} states, the Poisson log-normal distribution has an
upper tail asymptotically equal to the upper tail of the log-normal
distribution, so 
\begin{equation}
\sum_{y_{t}=M+1}^{\infty }\mathrm{PLN}\left( y_{t}|\mu _{t},\Sigma
_{tt}\right) \approx 1-\Phi \left( \dfrac{\log M-\mu _{t}}{\sqrt{\Sigma _{tt}%
}}\right) .  \label{PLN_tail_approximation}
\end{equation}%
Hence, we let the upper bound and lower bound to be 
\begin{eqnarray}
M_{U} &=&\left\lfloor \exp \left( z_{1-\alpha /2}\sqrt{\Sigma _{tt}}+\mu
_{t}\right) \right\rfloor +1,  \label{M_U} \\
M_{L} &=&\left\lfloor \exp \left( z_{\alpha /2}\sqrt{\Sigma _{tt}}+\mu
_{t}\right) \right\rfloor ,  \label{M_L}
\end{eqnarray}%
where $\left\lfloor x\right\rfloor $ means the integer part of a real number 
$x$ and $z_{\alpha }$ is the $\alpha $-quantile of a standard normal
distribution.

When $\alpha $ is small, we expect the true upper quantile for $\mathrm{PLN}%
\left( y_{t}|\mu _{t},\Sigma _{tt}\right) $ to be large, so the tail
approximation in equation (\ref{PLN_tail_approximation}) is particularly
accurate, and likewise for $M_{U}$ in equation (\ref{M_U}). On the other
hand, when $\mu _{t}$ is moderate, the true lower quantile is not far away
from $0$, which is usually the answer given by equation (\ref{M_L}). When $%
\mu _{t}$ is quite large, even the true lower quantile can be extreme; in
this case the tail approximation in equation (\ref{PLN_tail_approximation})
will become accurate again, so equation (\ref{M_L}) is still valid.

Simulation shows that both $M_{U}$ and $M_{L}$ offer practically useful
guidance for finding the upper and lower quantiles for $\mathrm{PLN}\left(
y_{t}|\mu _{t},\Sigma _{tt}\right) $, which then can help us to reduce the
total amount of calculations in equation (\ref{EIG_3}).

\section{Simulation\label{Simu.}}

In this Section we discuss two simulation examples, one using two
trigonometric templates to assess our algorithm and the other using three
real templates from astronomy. Here we only focus on comparing different
strategies for sequential design and demonstrating the faster speed of our
methodology. The comparison between our Bayesian methodology and the
existing frequentist inference methods is important but not our main focus
here. Besides, those existing methods employed in astronomy community does
not allow the on-line sequential learning, so neither do they have equal
status to compare with our method.

In the following examples, we only study three different types of strategies%
\footnote{%
One anonymous reader once suggested to add the comparison with GP-UCB
strategy proposed by \cite{Srinivas&Krause&Kakade&Seeger-2010} in our
simulation, but, as discussed in Section \ref{Sec.: literature}, their
setting is different from our work here and hence GP-UCB is not directly
applicable.}: the proposed sequential Monte Carlo strategy (SMCS) using
Algorithm \ref{Main-Alg}; the totally random strategy (TRS), by which the
choice of the filter is totally by chance; and the greedy strategy (GS),
which we deterministically choose the filters in the same order as the
absolute differences between the integrated intensities $\left\vert \Lambda
\left( B,\mu _{1}\right) -\Lambda \left( B,\mu _{2}\right) \right\vert $ for
different filter $B$ and two templates $\mu _{1},\mu _{2}$. Clearly, GS only
works when we have two template models. We note that TRS and GS are indeed
the current methodologies employed by astronomers.

This simulation section is primarily an illustration, for the real data
includes more domain-specific technicalities and will affect the clarity of
our presentation. These empirical results will be included in a follow-up
paper. Note that the outcomes and takeaways are similar for the real data as
for the simulation---SMCS will clearly perform better than the TRS, which is
commonly used in astronomy. Hence, the primary change in the real setting
will be to include a larger template base (over $100$ templates), but our
methodology can easily adapt to this much more complicate setting.

\paragraph{Example 1: Exponential Trigonometric Templates\label{ex1}}

We use two templates ($m=2$) and let%
\begin{eqnarray*}
\eta _{\mathrm{true}}\left( \nu \right) &=&\log \left( \lambda _{\mathrm{true%
}}\left( \nu \right) \right) =\omega _{1,\mathrm{ture}}\mu _{1}\left( \nu
\right) +\omega _{2,\mathrm{true}}\mu _{2}\left( \nu \right) ,\ \mathbf{%
\omega }_{\mathrm{true}}=\left( 0.8,0.2\right) ^{T}, \\
\mu _{1}\left( \nu \right) &=&2\sin \left( 2\pi \nu \right) +4,\ \mu
_{2}\left( \nu \right) =2\cos \left( 2\pi \nu \right) +4,\ \ \ \ \nu \in 
\left[ 0,1\right] , \\
k\left( \nu ,v^{\prime }\right) &=&\sigma ^{2}\exp \left( -\dfrac{\left( \nu
-\nu ^{\prime }\right) ^{2}}{2l^{2}}\right) ,\ \sigma =0.2,\ l=0.02,
\end{eqnarray*}%
and finally let the range of $y_{t}$ for calculating equation (\ref{EIG_3})
come from (\ref{M_U}) and (\ref{M_L}). The ten filters available here have
frequency range $\left[ 0,0.1\right] $, $\left[ 0.1,0.2\right] $,..., $\left[
0.9,1\right] $. For each strategy (TRS, GS, and SMCS), we run the simulation
up to $t=10$.

\begin{figure}[t]
\centering%
\begin{subfigure}[b]{0.48\textwidth}
  \centering
  \includegraphics[width=\linewidth]{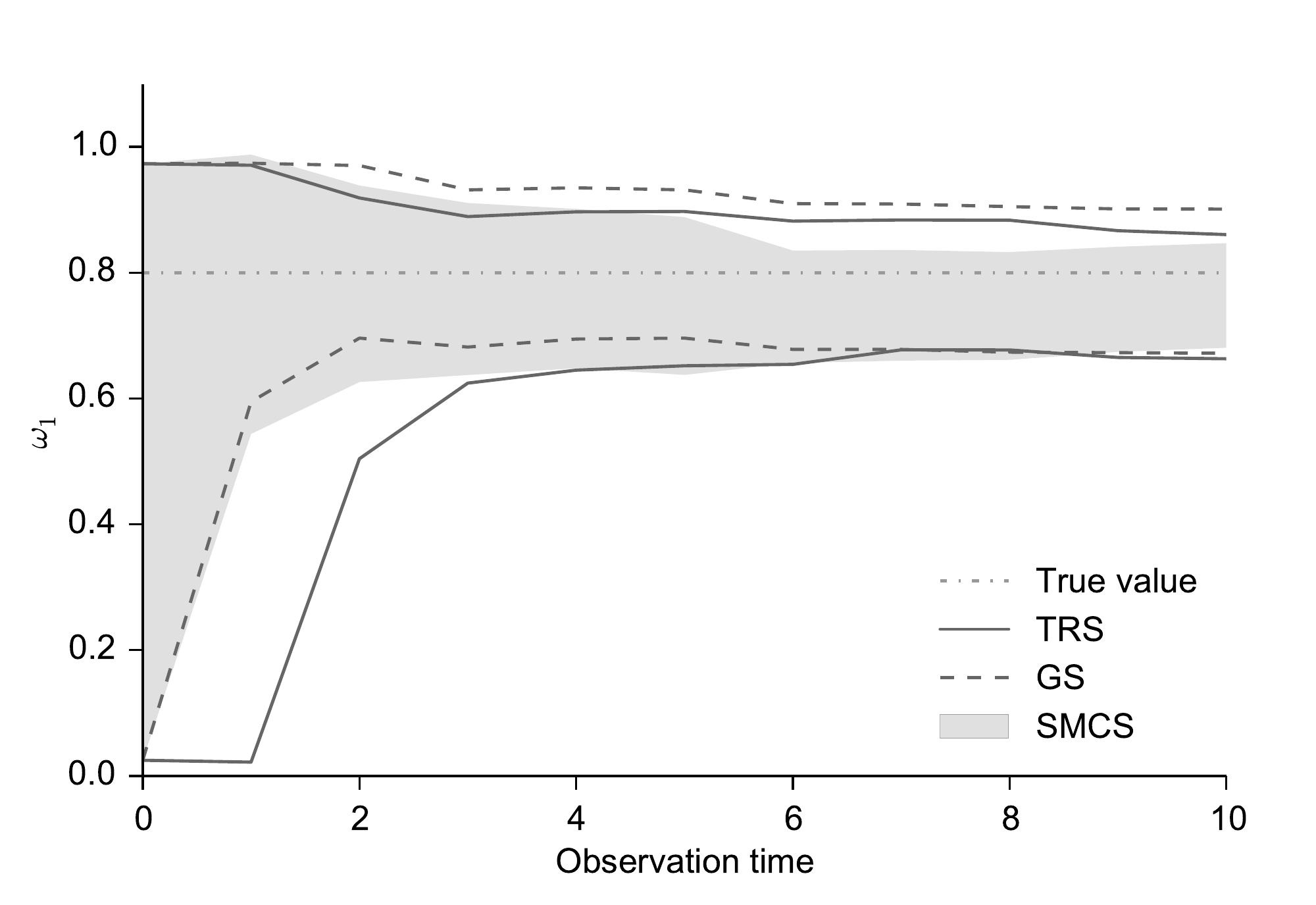}
  \caption{95\% posterior probability interval of $\omega_1$.}\label{sin_cos_example}
\end{subfigure}
\begin{subfigure}[b]{0.48\textwidth}
  \centering
  \includegraphics[width=\linewidth]{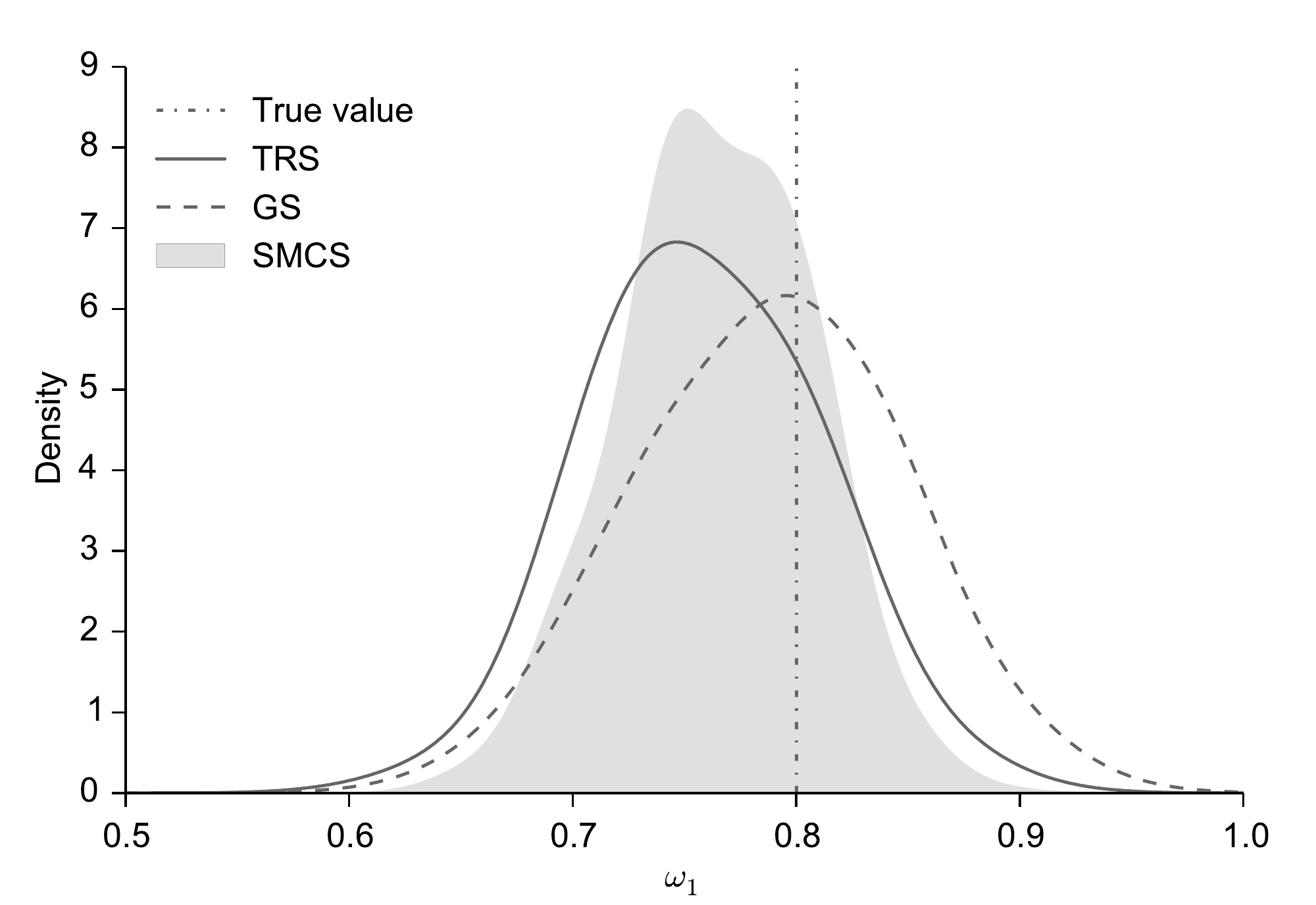}
  \caption{Posterior density of $\omega_1$ at $t=10$.}\label{ex1_posterior_density}
\end{subfigure}
\caption{A time-varying demonstration for the posterior distribution of $%
\protect\omega _{1}$ for Example 1.}
\end{figure}

The $95\%$ posterior probability intervals of $\omega _{1}$ for each time $t$
can be found in Figure \ref{sin_cos_example}. The probability interval of $%
\omega _{1}$ at $t=0$ comes from the uniform prior on $\left[ 0,1\right] $.
Both SMC and GS converge faster and give narrower intervals (since $t=1$)
than TRS. Figure \ref{ex1_posterior_density} also shows that both SMCS and
GS give a narrower and less biased result than TRS. In this case, SMCS
performs slightly better than GS in terms of the root of posterior mean
square error ($6.6\%$ for TRS, $6.0\%$ for GS, and $5.5\%$ for SMCS). Recall
that GS is valid only when $m=2$, but the proposed SMCS can be applied to
other cases. Actually, the generalization of GS for the case of $m>2$ is our
very first motivation to study this work.

\paragraph{Example 2: Active Galactic Nuclei (AGN), Composite (COMP), and
Starburst (SB) Templates\label{ex2}}

\begin{figure}[t]
\centering%
\begin{subfigure}[t]{0.48\textwidth}
    \centering
    \includegraphics[width=\linewidth]{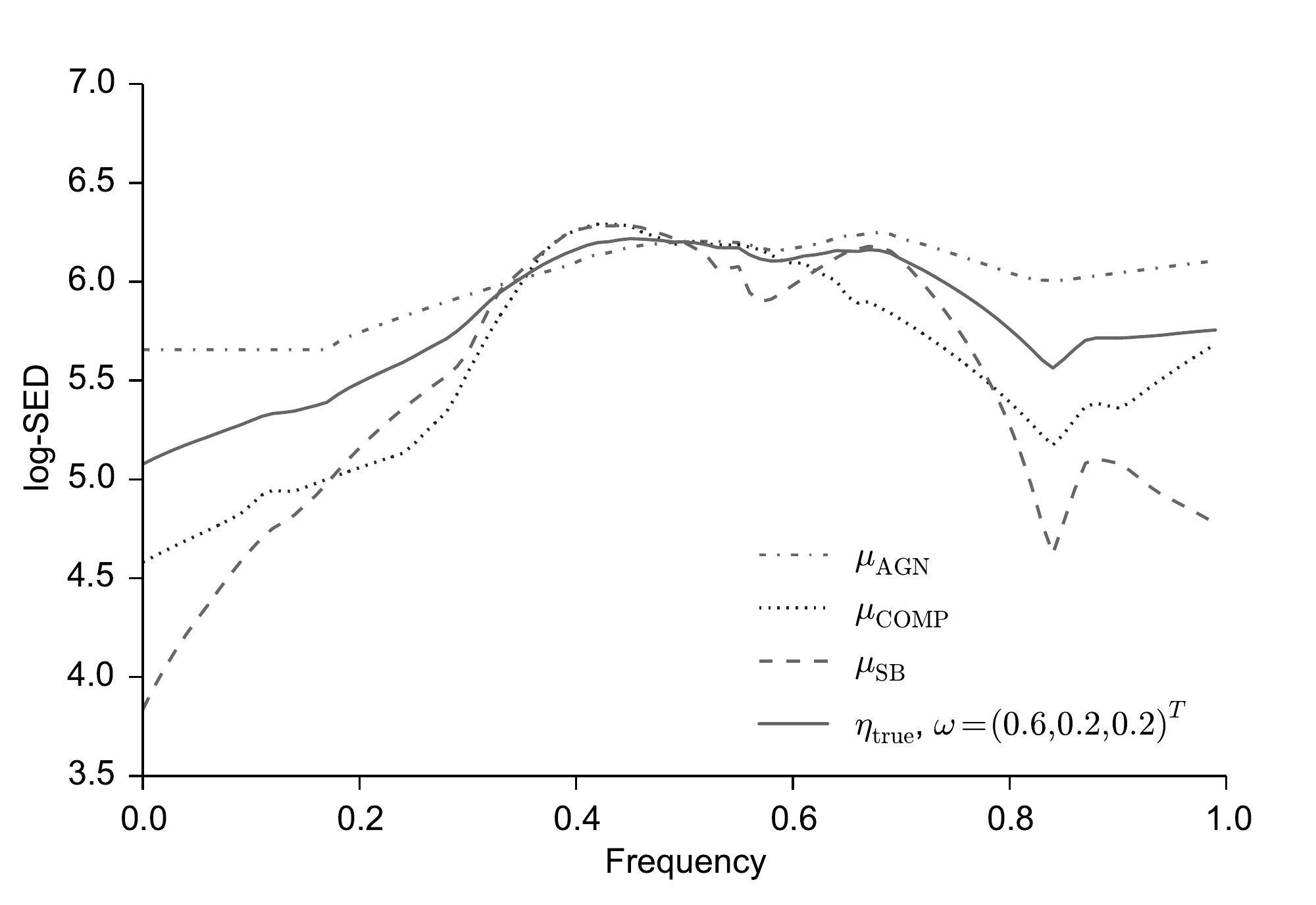}
    \caption{The three real templates from the astronomy. Also shown here is $\eta _{\mathrm{true}}$ we study in the Case 1.}
    \label{real_template}
\end{subfigure}
\begin{subfigure}[t]{0.48\textwidth}
    \centering
    \includegraphics[width=\linewidth]{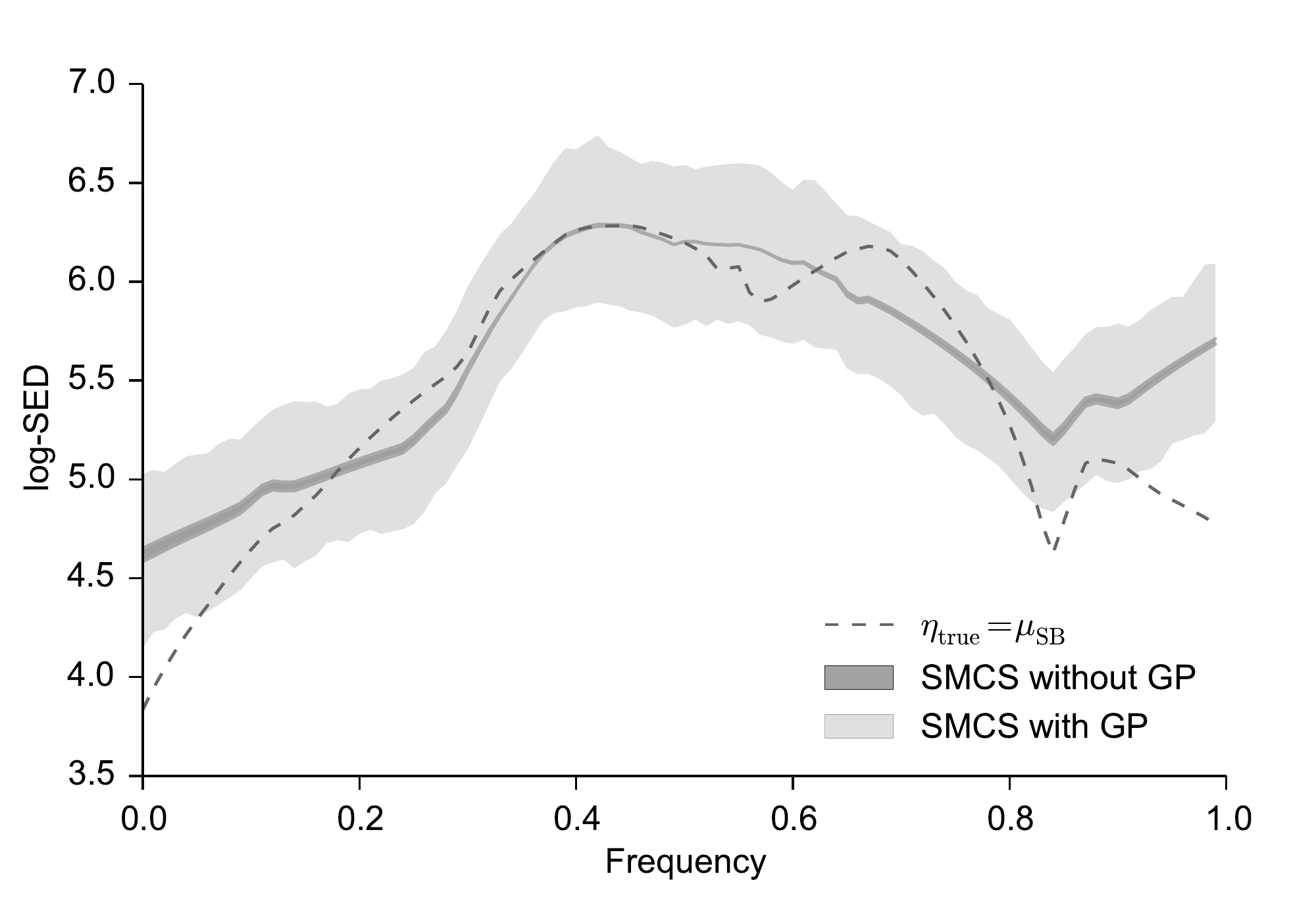}
    \caption{$95\%$ posterior region of $\protect\eta $ when $t=10$ in the Case 2.}
    \label{real_example}
\end{subfigure}
\caption{Illustrations of Example 2 in terms of log-SED's.}
\end{figure}

In this example, we set $m=3$ and use three real templates from astronomy:
AGN NGC5506, COMP IRAS 19254-7245, and SB NGC 7714 (\cite%
{Richards-2006,Elvis-1994,Hopkins-2007}). The frequency is scaled to $\left[
0,1\right] $ with the set of filters being the same as in Example 1. Now let%
\begin{equation*}
\eta _{\mathrm{true}}\left( \nu \right) =\omega _{1,\mathrm{ture}}\mu _{%
\mathrm{AGN}}\left( \nu \right) +\omega _{2,\mathrm{true}}\mu _{\mathrm{COMP}%
}\left( \nu \right) +\omega _{3,\mathrm{true}}\mu _{\mathrm{SB}}\left( \nu
\right) .
\end{equation*}%
The shape of $\mu _{\mathrm{AGN}}$, $\mu _{\mathrm{COMP}}$ and $\mu _{%
\mathrm{SB}}$ can be found in Figure \ref{real_template}. We consider two
cases to demonstrate the importance of including a Gaussian process (GP)
component in the modelling of the unknown $\eta \left( \nu \right) $.

\subparagraph{Case 1: Correctly specified templates.}

\begin{figure}[t]
\centering%
\begin{subfigure}[b]{0.48\textwidth}
  \centering
  \includegraphics[width=\linewidth]{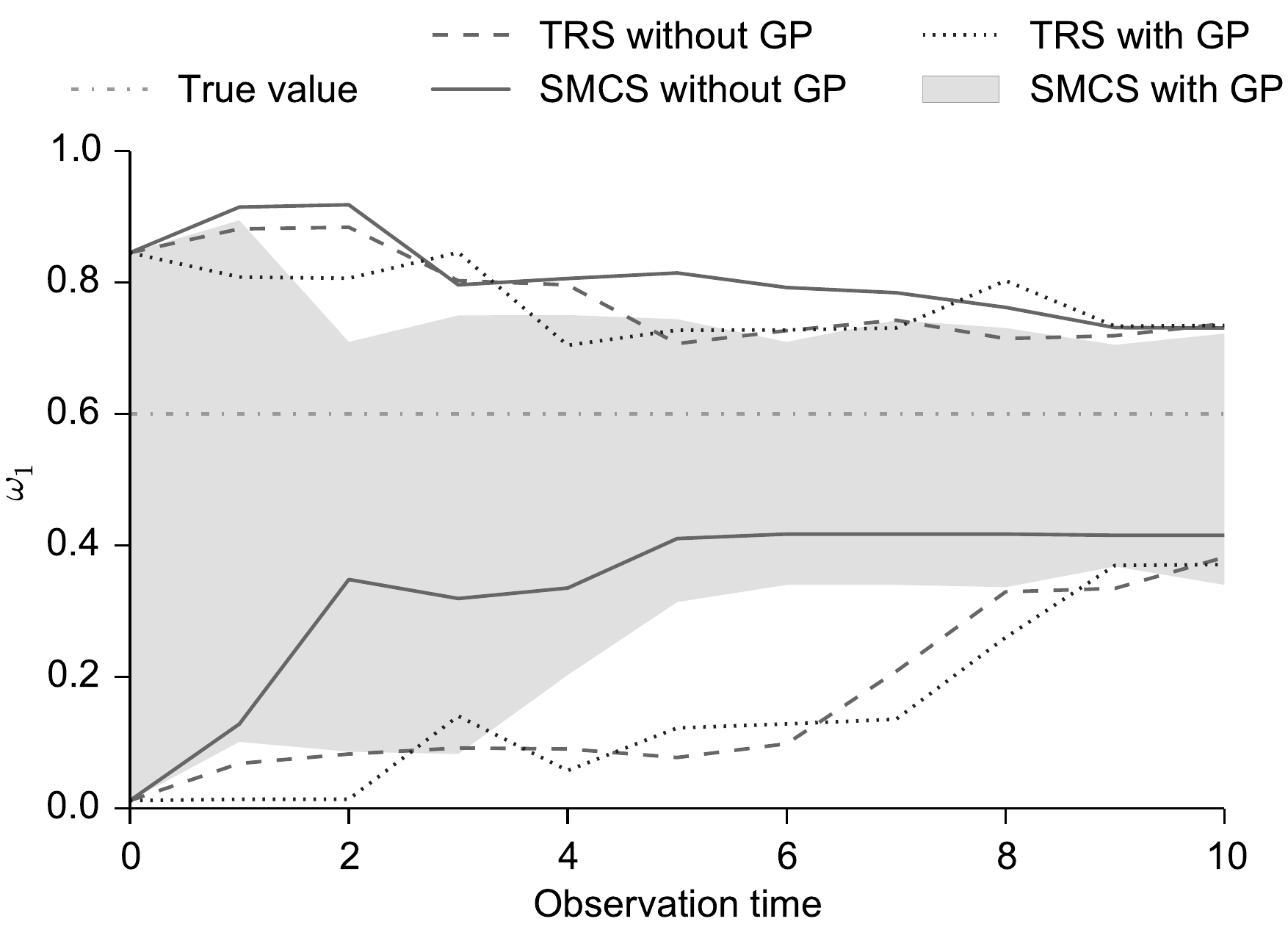}
  \caption{95\% posterior probability interval of $\omega_1$.}
\end{subfigure}
\begin{subfigure}[b]{0.48\textwidth}
  \centering
  \includegraphics[width=\linewidth]{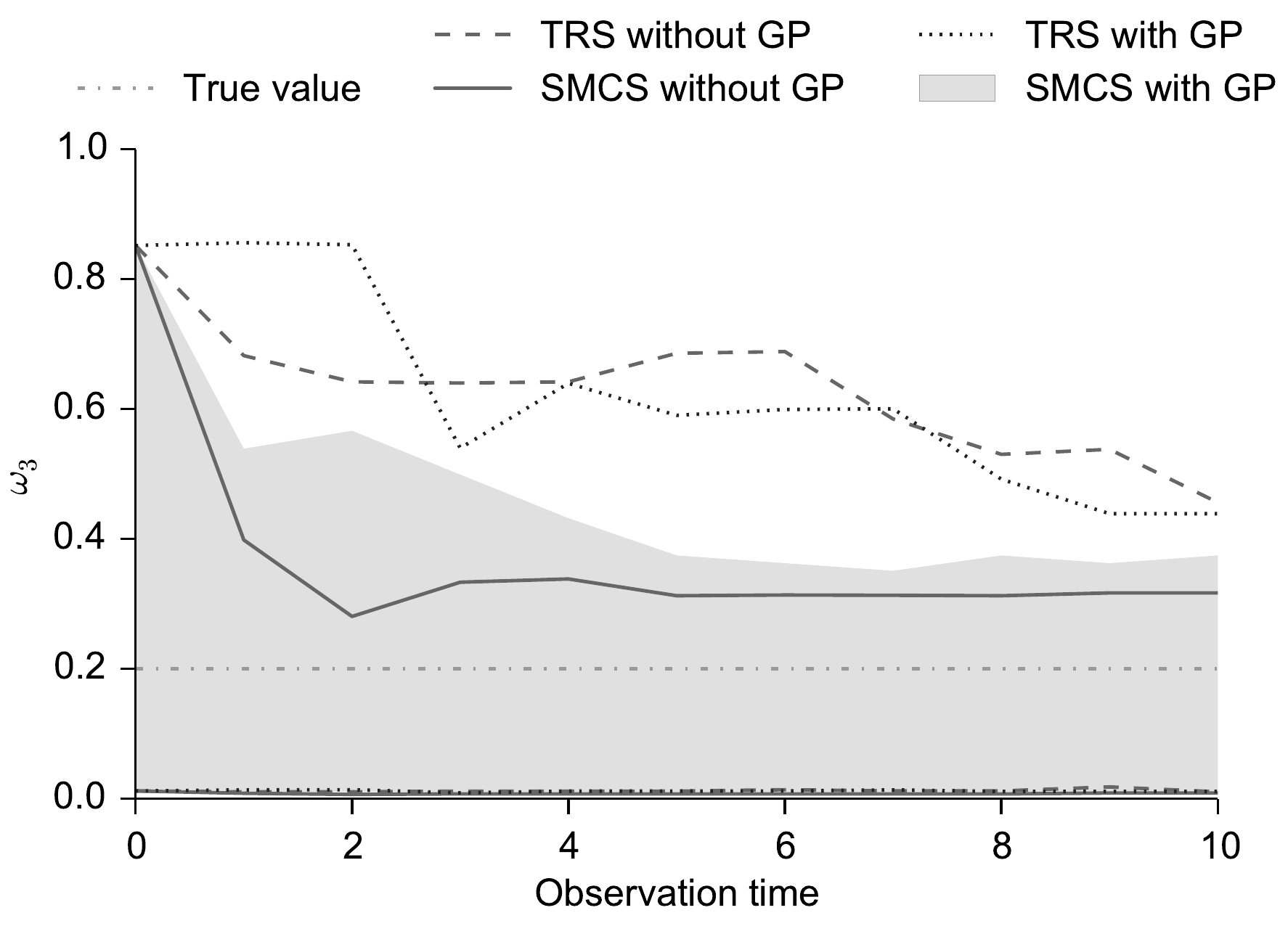}
  \caption{95\% posterior probability interval of $\omega_3$.}
\end{subfigure}
\caption{A time-varying demonstration for the posterior distribution of $%
\mathbf{\protect\omega }$ in the Case 1 of Example 2.}
\label{3_template_example}
\end{figure}

Let $\mathbf{\omega }_{\mathrm{true}}=\left( 0.6,0.2,0.2\right) ^{T}$ and
use all the three templates ($\mu _{\mathrm{AGN}}$, $\mu _{\mathrm{COMP}}$,
and $\mu _{\mathrm{SB}}$) for estimation. $\eta _{\mathrm{true}}$ in this
case is also plotted in Figure \ref{real_template}. We compare SMCS and TRS
under two scenarios: modelling $\eta $ with a GP prior and without a GP
prior. For the first scenario, we choose $k$ the same as in Example 1. The $%
95\%$ posterior probability intervals of $\omega _{1}$ and $\omega _{3}$ are
shown in Figure \ref{3_template_example}. SMCS without a GP prior results in
the narrowest interval because $\eta _{\mathrm{true}}$ is correctly
specified by the templates. On the other hand, no matter whether a GP prior
is used or not, SMCS converges faster than the TRS.

Here we emphasize that the slow convergence in Figure \ref%
{3_template_example} does not come from the approximation error of PoLNA (as
the two cases without GP does not require any approximation) but instead
from the small differences of the integrated intensities (for each filter)
among the templates. It seems that the three templates differ significantly
in Figure \ref{real_template}, but their actual integrated intensities do
not. 

\subparagraph{Case 2: Misspecified templates.}

Now we let $\mathbf{\omega }_{\mathrm{true}}=\left( 0,0,1\right) ^{T}$, so $%
\eta _{\mathrm{true}}=\mu _{\mathrm{SB}}$. In this case, we only use $\mu _{%
\mathrm{AGN}}$ and $\mu _{\mathrm{COMP}}$ for estimation, so any simple
convex combinations of $\mu _{\mathrm{AGN}}$ and $\mu _{\mathrm{COMP}}$ are
still far away from the truth. We compare the SMCS with and without a GP
prior using the same $k$ as in Example 1, and in Figure \ref{real_example}
we plot the $95\%$ posterior region of $\eta $ at $t=10$ (marginally for
each $\nu $). SMCS with a GP prior outperforms SMCS without a GP prior,
because the posterior region of the former is more capable of covering $\eta
_{\mathrm{true}}$. Precisely, with a GP prior on $\eta $ there is a $27\%$
posterior probability that 
\begin{equation*}
\left\Vert \eta -\eta _{\mathrm{true}}\right\Vert _{\infty }\leq \left\Vert
\mu _{\mathrm{COMP}}-\eta _{\mathrm{true}}\right\Vert _{\infty },
\end{equation*}%
where $\left\Vert f\right\Vert _{\infty }\triangleq \sup \left\{ \left\vert
f\left( \nu \right) \right\vert :\nu \in \left[ 0,1\right] \right\} $ and $%
\mu _{\mathrm{COMP}}$ is the most achievable estimation of $\eta $ without
using a GP prior. A GP prior on $\eta $ allows more adaptation to the data,
which leads to a more reliable conclusion---even though the templates we use
for estimation cannot completely describe the truth. This proves the
potential and the practical value of our methodology.

\section{Discussion\label{Conc.}}

\subsection{Differences with related works\label{Sec.: literature}}

\subparagraph{Bayesian experimental design versus Bayesian optimization.}

The use of Gaussian process (GP) to sequentially choose the next experiment
to perform so as to optimize some value of information can be traced back to 
\cite{Kushner-1963}. The research alone this line is generally called \emph{%
Bayesian optimization}. More literature reviews in this direction can be
found in the survey papers \cite{Brochu&Cora&deFreitas-2010} and \cite%
{Frazier-2010}. In this context, \cite{Villemonteix&Vazquez&Walter-2009}
also uses an information theoretic criterion similar to our work.

However, our formulation of Bayesian sequential experimental design is
fundamentally different from Bayesian optimization. Optimization aims to
find the optimum of a given target (a local criterion); experimental design
seeks to characterize the entire target distribution (a global criterion),
as measured by some (set of) aggregate information metric(s). Hence, the
focus of our approach is not the same as those Bayesian optimization
literatures.

\subparagraph{Sequential design versus non-sequential design.}

There are indeed some literatures study experimental design with global
criterion using the Bayesian nonparametric GP prior. For example, \cite%
{Sacks&Schiller&Welch-1989} seeks to minimize integrated mean square error,
while \cite{Shewry&Wynn-1987} and \cite%
{Currin&Mitchell&Morris&Ylvisaker-1991} want to maximize the entropy of the
posterior. However, these papers do not study the sequential procedure, so
they are different from our paper.

As far as we know, only few existing papers aim to \emph{globally}
understand a target, studying \emph{sequential} design and \emph{Bayesian
nonparametric} estimation at the same time, which is what we have done in
this paper. The most matchable literature we know so far regarded to the
intersection between Bayesian nonparametrics and Bayesian sequential
experimental design is \cite{Ferreira&Sanyal-2014}, which still differs a
lot from our work not only in the scientific goal, in the design problem
formulation, but also in the inferential techniques.

\subparagraph{Non-conjugate model versus Gaussian conjugate model.}

Even though \cite{Ferreira&Sanyal-2014} is the most similar literature to
our work so far, they consider only the \emph{conjugate} GP model with
continuous Gaussian measurement error for the observations. Actually, most
of the existing GP literatures (whether focusing on Bayesian optimization,
active learning, or other problems) assume the Gaussian conjugacy in their
model setting, in which case the inference techniques are still based on
closed analytical forms.

However, our approach is capable of dealing with \emph{non-conjugate}
Bayesian nonparametric models. Our paper not only adopts a non-conjugate
normal-Poisson model but also provides an efficient inference technique
(SMC) that can generalize to other nonparametric priors. Even though \cite%
{Gramacy&Polsona-2011} also study SMC inference for the sequential design on
a GP, they neither focus on a global criterion to understand the target nor
use a non-conjugate model.

\subparagraph{Design versus inference in the non-conjugate hierarchical
Bayesian nonparametric model.}

Our paper studies sequential active learning in addition to performing
inference in a non-conjugate hierarchical Bayesian nonparametric model.
While many papers have studied the inference portion of this task (usually
based on LGCP, such as \cite{Rue&Martino-2009} and \cite%
{Simpson&Illian&Lindgren&Sorbye&Rue-2013}), none have simultaneously tackled
the design problem, largely due to the huge computational cost. In this
paper, we have created an efficient computational approach to solve this
problem, demonstrating its usefulness on an astronomical application.

\subsection{Conclusion}

In this paper, we first study the problem of Bayesian nonparametric
sequential experimental design (BNS-ED) without conjugacy. We build a
three-level hierarchal Bayesian nonparametric model that aims to find the
optimal astronomical observational strategy. A sequential Monte Carlo (SMC)
strategy is then proposed to solve the problem of interest. To overcome the
computation hurdle inherited naturally from the log-Gaussian Cox process
(LGCP), we exploit the special features of this model and provide a new
inference technique for it, called Poisson log-normal approximation (PoLNA),
which can still be applied even for spatial-temporal LGCP.

We would like to emphasize that even though we mainly focus on the widely
studied LGCP in this paper, the BNS-ED framework we discuss here and the
corresponding inferential scheme, which employs sequential Monte Carlo
techniques, can be easily generalized to other nonparametric models.

The computation problem encountered in this paper is generally difficult, so
we suspect that this is why there is no existing work on this
topic---sequential design for globally learning a nonparametric target with
non-conjugate sampling distribution. Our computational technique provides
the algorithmic speed to make this idea practical on real applications,
which then justifies the novelty and value of our approach.

{\small 
\bibliographystyle{abbrvnat}
\bibliography{../ju-chen,../Pavlos,../NIPS_Reviewer}
} \appendix

\section{Derivations\label{Derivations}}

\subsection{Equation (\protect\ref{Bayesian_Updating})}

By Bayes Theorem,%
\begin{eqnarray*}
p\left( \mathbf{\omega }|y_{t},B_{t},\mathcal{F}_{t-1}\right) &=&\dfrac{%
p\left( y_{t},B_{t}|\mathbf{\omega },\mathcal{F}_{t-1}\right) p\left( 
\mathbf{\omega }|\mathcal{F}_{t-1}\right) }{p\left( y_{t},B_{t}|\mathcal{F}%
_{t-1}\right) }=\dfrac{p\left( y_{t}|B_{t},\mathbf{\omega },\mathcal{F}%
_{t-1}\right) p\left( B_{t}|\mathbf{\omega },\mathcal{F}_{t-1}\right)
p\left( \mathbf{\omega }|\mathcal{F}_{t-1}\right) }{p\left( y_{t},B_{t}|%
\mathcal{F}_{t-1}\right) } \\
&=&\dfrac{p\left( y_{t}|B_{t},\mathbf{\omega },\mathcal{F}_{t-1}\right)
p\left( B_{t}|\mathcal{F}_{t-1}\right) p\left( \mathbf{\omega }|\mathcal{F}%
_{t-1}\right) }{p\left( y_{t},B_{t}|\mathcal{F}_{t-1}\right) }=\dfrac{%
p\left( y_{t}|B_{t},\mathbf{\omega },\mathcal{F}_{t-1}\right) }{p\left(
y_{t}|B_{t},\mathcal{F}_{t-1}\right) }p\left( \mathbf{\omega }|\mathcal{F}%
_{t-1}\right) ,
\end{eqnarray*}%
where the third line follows from $p\left( B_{t}|\mathbf{\omega },\mathcal{F}%
_{t-1}\right) =p\left( B_{t}|\mathcal{F}_{t-1}\right) $ because by
definition $B_{t}$ and $\mathbf{\omega }$ are independent given $\mathcal{F}%
_{t-1}$.

\subsection{Equation (\protect\ref{EIG_2})}

Let us simplify the expected information gain $EIG_{t}\left( B_{t}\right) $
in equation (\ref{EIG}). We note that%
\begin{eqnarray*}
D_{\mathrm{KL}}\left( p\left( \mathbf{\omega }|y_{t},B_{t},\mathcal{F}%
_{t-1}\right) ||p\left( \mathbf{\omega }|\mathcal{F}_{t-1}\right) \right)
&=&\int \log \left( \dfrac{p\left( \mathbf{\omega }|y_{t},B_{t},\mathcal{F}%
_{t-1}\right) }{p\left( \mathbf{\omega }|\mathcal{F}_{t-1}\right) }\right)
p\left( \mathbf{\omega }|y_{t},B_{t},\mathcal{F}_{t-1}\right) \mathrm{d}%
\mathbf{\omega } \\
&=&\int \log \left( \dfrac{p\left( y_{t}|B_{t},\mathbf{\omega },\mathcal{F}%
_{t-1}\right) }{p\left( y_{t}|B_{t},\mathcal{F}_{t-1}\right) }\right) \dfrac{%
p\left( y_{t}|B_{t},\mathbf{\omega },\mathcal{F}_{t-1}\right) }{p\left(
y_{t}|B_{t},\mathcal{F}_{t-1}\right) }p\left( \mathbf{\omega }|\mathcal{F}%
_{t-1}\right) \mathrm{d}\mathbf{\omega },
\end{eqnarray*}%
where the second line follows from equation (\ref{Bayesian_Updating}).
Therefore, by swapping the integral and the summation,%
\begin{eqnarray*}
EIG_{t}\left( B_{t}\right) &=&\sum_{y_{t}=0}^{\infty }D_{\mathrm{KL}}\left(
p\left( \mathbf{\omega }|y_{t},B_{t},\mathcal{F}_{t-1}\right) ||p\left( 
\mathbf{\omega }|\mathcal{F}_{t-1}\right) \right) p\left( y_{t}|B_{t},%
\mathcal{F}_{t-1}\right) \\
&=&\int \sum_{y_{t}=0}^{\infty }\log \left( \dfrac{p\left( y_{t}|B_{t},%
\mathbf{\omega },\mathcal{F}_{t-1}\right) }{p\left( y_{t}|B_{t},\mathcal{F}%
_{t-1}\right) }\right) p\left( y_{t}|B_{t},\mathbf{\omega },\mathcal{F}%
_{t-1}\right) p\left( \mathbf{\omega }|\mathcal{F}_{t-1}\right) \mathrm{d}%
\mathbf{\omega } \\
&=&\mathbb{E}_{\mathbf{\omega }|\mathcal{F}_{t-1}}\left( D_{KL}\left(
p\left( y_{t}|B_{t},\mathbf{\omega },\mathcal{F}_{t-1}\right) ||p\left(
y_{t}|B_{t},\mathcal{F}_{t-1}\right) \right) \right) .
\end{eqnarray*}

\subsection{Equation (\protect\ref{marginal_like})}

Recall that%
\begin{eqnarray*}
p\left( y_{t}|B_{t},\mathcal{F}_{t-1},\mathbf{\omega }\right) &=&\dfrac{%
p\left( y_{t},B_{t}|\mathcal{F}_{t-1},\mathbf{\omega }\right) }{p\left(
B_{t}|\mathcal{F}_{t-1},\mathbf{\omega }\right) }=\dfrac{\int p\left(
y_{t},B_{t}|\mathcal{F}_{t-1},\eta ,\mathbf{\omega }\right) p\left( \eta |%
\mathcal{F}_{t-1},\mathbf{\omega }\right) \mathrm{d}\eta }{p\left( B_{t}|%
\mathcal{F}_{t-1},\mathbf{\omega }\right) } \\
&=&\dfrac{\dint \left( p\left( y_{t}|B_{t},\mathcal{F}_{t-1},\eta ,\mathbf{%
\omega }\right) p\left( B_{t}|\mathcal{F}_{t-1},\eta ,\mathbf{\omega }%
\right) p\left( \eta |\mathcal{F}_{t-1},\mathbf{\omega }\right) \right) 
\mathrm{d}\eta }{p\left( B_{t}|\mathcal{F}_{t-1},\mathbf{\omega }\right) } \\
&=&\int p\left( y_{t}|B_{t},\eta \right) p\left( \eta |\mathcal{F}_{t-1},%
\mathbf{\omega }\right) \mathrm{d}\eta ,
\end{eqnarray*}%
and%
\begin{eqnarray*}
p\left( \eta |\mathcal{F}_{t-1},\mathbf{\omega }\right) &=&\dfrac{p\left(
y_{t-1},B_{t-1}|\eta ,\mathcal{F}_{t-2},\mathbf{\omega }\right) p\left( \eta
|\mathcal{F}_{t-2},\mathbf{\omega }\right) }{p\left( y_{t-1},B_{t-1}|%
\mathcal{F}_{t-2},\mathbf{\omega }\right) } \\
&=&\dfrac{p\left( y_{t-1}|B_{t-1},\eta ,\mathcal{F}_{t-2},\mathbf{\omega }%
\right) p\left( B_{t-1}|\mathcal{F}_{t-2},\mathbf{\omega }\right) p\left(
\eta |\mathcal{F}_{t-2},\mathbf{\omega }\right) }{p\left( y_{t-1}|B_{t-1},%
\mathcal{F}_{t-2},\mathbf{\omega }\right) p\left( B_{t-1}|\mathcal{F}_{t-2},%
\mathbf{\omega }\right) } \\
&=&\dfrac{p\left( y_{t-1}|B_{t-1},\eta \right) p\left( \eta |\mathcal{F}%
_{t-2},\mathbf{\omega }\right) }{p\left( y_{t-1}|B_{t-1},\mathcal{F}_{t-2},%
\mathbf{\omega }\right) }=\cdots =\dfrac{\prod_{s=1}^{t-1}p\left(
y_{s}|B_{s},\eta \right) }{\prod_{s=1}^{t-1}p\left( y_{s}|B_{s},\mathcal{F}%
_{s-1},\mathbf{\omega }\right) }p\left( \eta |\mathbf{\omega }\right) .
\end{eqnarray*}%
Combining the two formulas above will give us equation (\ref{marginal_like}).

\subsection{Equation (\protect\ref{PLN_as_Laplace_Trans})}

We have%
\begin{eqnarray*}
\mathrm{PLN}\left( \mathbf{y}|\mathbf{\tilde{\mu}},\mathbf{\tilde{\Sigma}}%
\right)  &=&\int \left( \prod_{s=1}^{u}\prod_{n=1}^{n_{s}}\mathrm{Poisson}%
\left( \tilde{y}_{sn}\left\vert \tilde{\Lambda}_{s}\right. \right) \mathcal{N%
}_{u}\left( \left. \log \mathbf{\tilde{\Lambda}}\right\vert \mathbf{\tilde{%
\mu}},\mathbf{\tilde{\Sigma}}\right) \right) \mathrm{d}\left( \log \mathbf{%
\tilde{\Lambda}}\right)  \\
&=&\dfrac{\det \left( 2\pi \mathbf{\tilde{\Sigma}}\right) ^{-1/2}}{%
\prod_{s=1}^{n}\prod_{n=1}^{n_{s}}\tilde{y}_{sn}!}\int e^{-\mathbf{n}^{T}e^{%
\mathbf{x}}+\mathbf{S}^{T}\mathbf{x}}e^{-\left( \mathbf{x}-\mathbf{\tilde{\mu%
}}\right) ^{T}\mathbf{\tilde{\Sigma}}^{-1}\left( \mathbf{x}-\mathbf{\tilde{%
\mu}}\right) /2}\mathrm{d}\mathbf{x} \\
&=&\dfrac{\det \left( 2\pi \mathbf{\tilde{\Sigma}}\right) ^{-1/2}e^{\left( 
\mathbf{\tilde{\mu}}+\mathbf{\tilde{\Sigma}S}\right) ^{T}\mathbf{\tilde{%
\Sigma}}^{-1}\left( \mathbf{\tilde{\mu}}+\mathbf{\tilde{\Sigma}S}\right) /2-%
\mathbf{\tilde{\mu}}^{T}\mathbf{\tilde{\Sigma}}^{-1}\mathbf{\tilde{\mu}}/2}}{%
\prod_{s=1}^{n}\prod_{n=1}^{n_{s}}\tilde{y}_{sn}!} \\
&&\times \int e^{-\mathbf{n}^{T}e^{\mathbf{x}}}e^{-\left( \mathbf{x}-\mathbf{%
\tilde{\mu}}-\mathbf{\tilde{\Sigma}S}\right) ^{T}\mathbf{\tilde{\Sigma}}%
^{-1}\left( \mathbf{x}-\mathbf{\tilde{\mu}}-\mathbf{\tilde{\Sigma}S}\right)
/2}\mathrm{d}\mathbf{x} \\
&\mathbf{=}&\dfrac{e^{\mathbf{S}^{T}\mathbf{\tilde{\Sigma}S}/2+\mathbf{%
\tilde{\mu}}^{T}\mathbf{S}}}{\prod_{s=1}^{n}\prod_{n=1}^{n_{s}}\tilde{y}%
_{sn}!}\int e^{-\mathbf{n}^{T}\mathbf{\tilde{\Lambda}}}\log \mathcal{N}%
_{u}\left( \left. \mathbf{\tilde{\Lambda}}\right\vert \mathbf{\tilde{\mu}}+%
\mathbf{\tilde{\Sigma}S},\mathbf{\tilde{\Sigma}}\right) \mathrm{d}\mathbf{%
\tilde{\Lambda}},
\end{eqnarray*}%
where $\tilde{\Lambda}_{s}$'s are the unique components of $\mathbf{\Lambda }
$ with corresponding multivariate log-normal parameters being $\mathbf{%
\tilde{\mu}}$ and $\mathbf{\tilde{\Sigma}}$ and%
\begin{eqnarray*}
\tilde{y}_{sn}|\tilde{\Lambda}_{s} &\backsim &\mathrm{Poisson}\left( \tilde{%
\Lambda}_{s}\right) \text{ i.i.d. for all }n=1,...,n_{s},\ s=1,...,u; \\
\mathbf{y} &=&\left. \left[ \tilde{y}_{sn}\right] _{n=1}^{n_{s}}\right.
_{s=1}^{u};\ \mathbf{S}\triangleq \left( \sum_{n=1}^{n_{1}}\tilde{y}%
_{1n},...,\sum_{n=1}^{n_{u}}\tilde{y}_{un}\right) ^{T}; \\
\mathbf{n} &\triangleq &\left( n_{1},...,n_{u}\right) ^{T};\ \mathbf{x}%
\triangleq \left( \log \tilde{\Lambda}_{1},...,\log \tilde{\Lambda}%
_{u}\right) ^{T}.
\end{eqnarray*}

\section{PoLNA Parameters Calculation\label{get_par_multi}}

In this Section, we will introduce the detailed procedures to calculate $%
\mathbf{\mu }_{\Lambda }$ and $\mathbf{\Sigma }_{\Lambda }$. After a fine
discretization on the frequency space ($\nu $), we will have 
\begin{eqnarray*}
\mathbf{\eta } &\sim &\mathcal{N}_{D}\left( \mathbf{\mu }_{\eta },\mathbf{K}%
\right) ,\ \left[ \mathbf{\eta }\right] _{j}\triangleq \eta \left( \nu
^{\left( j\right) }\right) , \\
\left[ \mathbf{\mu }_{\eta }\right] _{j} &\triangleq &\sum_{i=1}^{m}\omega
_{i}\mu _{i}\left( \nu ^{\left( j\right) }\right) ,\ \left[ \mathbf{K}\right]
_{j,j^{\prime }}\triangleq \left[ k\left( \nu ^{\left( j^{\prime }\right)
},\nu ^{\left( j^{\prime }\right) }\right) \right] ,\ \ \ \ j,j^{\prime
}=1,...,D, \\
\Lambda \left( B,\eta \right) &\approx &\sum_{\nu ^{\left( j\right) }\in
B}e^{\eta \left( \nu ^{\left( j\right) }\right) }\left( \nu ^{\left(
j\right) }-\nu ^{\left( j-1\right) }\right) .
\end{eqnarray*}%
$D$ is the number of discretization.

The distribution of $\Lambda \left( B,\eta \right) $ is approximated by a
log-normal distribution. We have 
\begin{equation*}
\log \left( \Lambda \left( B,\eta \right) \right) =\log \left( \sum_{\nu
^{\left( j\right) }\in B}e^{\eta \left( \nu ^{\left( j\right) }\right)
}\left( \nu ^{\left( j\right) }-\nu ^{\left( j-1\right) }\right) \right)
=\log \left( \sum_{\nu ^{\left( j\right) }\in B}e^{\eta \left( \nu ^{\left(
j\right) }\right) +\log \left( \nu ^{\left( j\right) }-\nu ^{\left(
j-1\right) }\right) }\right) .
\end{equation*}%
Also, $\eta \left( \nu ^{\left( j\right) }\right) +\log \left( \nu ^{\left(
j\right) }-\nu ^{\left( j-1\right) }\right) $ follows a multivariate normal
distribution. Our problem now is equal to the following one: 
\begin{equation*}
\left( y_{1}^{1},\ldots ,y_{n_{1}}^{1},\ldots
,y_{1}^{m},y_{n_{m}}^{m}\right) \sim \mathcal{N}_{n}\left( \mathbf{\mu }_{y},%
\mathbf{\Sigma }_{y}\right)
\end{equation*}%
Here, $n=\sum_{i=1}^{m}n_{i}$. Let $s^{i}=\log \left(
\sum_{j=1}^{n_{i}}e^{y_{j}^{i}}\right) $. Approximately, 
\begin{equation*}
\left( s^{1},\ldots ,s^{m}\right) \sim \mathcal{N}_{m}\left( \mathbf{\mu }%
_{s},\mathbf{\Sigma }_{s}\right)
\end{equation*}

For all $i,1\leq k\leq n_{i}$, define $s_{k}^{i}=\sum_{j=1}^{k}y_{k}^{i}$,
so $s_{n_{i}}^{i}=s^{i}$. Then 
\begin{equation*}
s_{k}^{i}=\log \left( e^{s_{k-1}^{i}}+e^{y_{k}^{i}}\right) =s_{k-1}^{i}+\log
\left( 1+e^{w_{k}^{i}}\right)
\end{equation*}%
where $w_{k}^{i}=y_{k}^{i}-s_{k-1}^{i}$.

For any two random variables $X$ and $Y$, let $\mu _{X}\triangleq \mathbb{E}%
\left( X\right) $, $\sigma _{X}^{2}\triangleq \func{Var}\left( X\right) $,
and $\rho _{XY}\triangleq \func{Cor}\left( X,Y\right) $. Following \cite%
{Safak-1993}, we have 
\begin{eqnarray*}
\mu _{s_{k}^{i}} &=&\mu _{s_{k-1}^{i}}+G_{1}\left( \sigma _{w_{k}^{i}},\mu
_{w_{k}^{i}}\right) \\
\mu _{w_{k}^{i}} &=&\mu _{y_{k}^{i}}-\mu _{s_{k-1}^{i}} \\
\sigma _{w_{k}^{i}}^{2} &=&\sigma _{y_{k}^{i}}^{2}-\sigma
_{s_{k-1}^{i}}^{2}-2\rho _{s_{k-1}^{i},y_{k}^{i}}\sigma _{y_{k}^{i}}\sigma
_{s_{k-1}^{i}} \\
\sigma _{s_{k}^{i}}^{2} &=&\sigma _{s_{k-1}^{i}}^{2}-G_{1}^{2}\left( \sigma
_{w_{k}^{i}},\mu _{w_{k}^{i}}\right) +G_{2}\left( \sigma _{w_{k}^{i}},\mu
_{w_{k}^{i}}\right) +2\dfrac{\sigma _{s_{k-1}^{i}}}{\sigma _{w_{k}^{i}}^{2}}%
\left( \rho _{s_{k-1}^{i},y_{k}^{i}}\sigma _{y_{k}^{i}}-\sigma
_{s_{k-1}^{i}}\right) G_{3}\left( \sigma _{w_{k}^{i}},\mu _{w_{k}^{i}}\right)
\\
\rho _{s_{j}^{i},y_{k}^{i}} &=&\rho _{s_{j-1}^{i},y_{k}^{i}}\dfrac{\sigma
_{s_{j-1}^{i}}}{\sigma _{s_{j}^{i}}}\left( 1-\dfrac{G_{3}\left( \sigma
_{w_{j}^{i}},\mu _{w_{j}^{i}}\right) }{\sigma _{w_{j}^{i}}^{2}}\right) +\rho
_{y_{j}^{i},y_{k}^{i}}\dfrac{\sigma _{y_{j}^{i}}}{\sigma _{s_{j}^{i}}\sigma
_{w_{j}^{i}}^{2}}\dfrac{G_{3}\left( \sigma _{w_{j}^{i}},\mu
_{w_{j}^{i}}\right) }{\sigma _{w_{j}^{i}}^{2}},
\end{eqnarray*}%
where $G1$, $G2$, and $G3$ will be defined later.

We have two methods to compute $\rho _{s_{n_{i}}^{i},s_{n_{j}}^{j}}$. For
any $i,j$ and $1\leq k_{1}\leq n_{i}-1$, $1\leq k_{2}\leq n_{j}-1$, suppose
we already know $\rho _{s_{k_{1}}^{i},s_{k_{2}}^{j}}$. Then we could first
update $\rho _{s_{k_{1}+1}^{i},s_{k_{2}}^{j}}$ and then $\rho
_{s_{k_{1}+1}^{i},s_{k_{2}+2}^{j}}$. The strategy is as follows:%
\begin{equation*}
\mathbb{E}\left( \left. s_{k_{2}}^{j}-\mu _{s_{k_{2}}^{j}}\right\vert
w_{k_{1}+1}^{i}\right) =\dfrac{\rho _{w_{k_{1}+1}^{i},s_{k_{2}}^{j}}\sigma
_{s_{k_{2}}^{j}}}{\sigma _{w_{k_{1}+1}^{i}}}\left( w_{k_{1}+1}^{i}-\mu
_{w_{k_{1}+1}^{i}}\right) ,
\end{equation*}%
where%
\begin{eqnarray*}
\rho _{s_{k_{1}+1}^{i},s_{k_{2}}^{j}} &=&\dfrac{\mathbb{E}\left( \left(
s_{k_{1}}^{i}-\mu _{s_{k_{1}}^{i}}+\log \left( 1+e^{w_{k_{1}+1}^{i}}\right)
\right) \left( s_{k_{2}}^{j}-\mu _{s_{k_{2}}^{j}}\right) \right) }{\sigma
_{s_{k_{1}+1}^{i}}\sigma _{s_{k_{2}}^{j}}} \\
&=&\rho _{s_{k_{1}}^{i},s_{k_{2}}^{j}}\dfrac{\sigma _{s_{k_{1}}^{i}}}{\sigma
_{s_{k_{1}+1}^{i}}}+\rho _{w_{k_{1}+1}^{i},s_{k_{2}}^{j}}\dfrac{G_{3}\left(
\sigma _{w_{k_{1}+1}^{i}},\mu _{w_{k_{1}+1}^{i}}\right) }{\sigma
_{s_{k_{1}+1}^{i}}\sigma _{w_{k_{1}+1}^{i}}}.
\end{eqnarray*}%
We could have similar formulas for $\rho _{s_{k_{1}+1}^{i},s_{k_{2}+2}^{j}}$.

For a special case, if we have $n_i=n_j$ and already know $%
\rho_{s^i_k,s^j_k} $, $1\leq k \leq n_i-1=n_j-1$, we will have

\begin{eqnarray*}
\rho _{s_{k+1}^{i},s_{k+1}^{j}} &=&\dfrac{\mathbb{E}\left( \left( 
\begin{array}{c}
s_{k}^{i}-\mu _{s_{k}^{i}}+\log \left( 1+e^{w_{k+1}^{i}}\right) \\ 
-G_{1}\left( \sigma _{w_{k+1}^{i}},\mu _{w_{k+1}^{i}}\right)%
\end{array}%
\right) \left( 
\begin{array}{c}
s_{k}^{j}-\mu _{s_{k}^{j}}+\log \left( 1+e^{w_{k+1}^{j}}\right) \\ 
-G_{1}\left( \sigma _{w_{k+1}^{j}},\mu _{w_{k+1}^{j}}\right)%
\end{array}%
\right) \right) }{\sigma _{s_{k+1}^{i}}\sigma _{s_{k+1}^{j}}} \\
&=&\rho _{s_{k}^{i},s_{k}^{j}}\dfrac{\sigma _{s_{k}^{i}}\sigma _{s_{k}^{j}}}{%
\sigma _{s_{k+1}^{i}}\sigma _{s_{k+1}^{j}}}+\rho _{w_{k+1}^{i},s_{k}^{j}}%
\dfrac{\sigma _{s_{k}^{j}}G_{3}\left( \sigma _{w_{k+1}^{i}},\mu
_{w_{k+1}^{i}}\right) }{\sigma _{s_{k+1}^{i}}\sigma _{s_{k+1}^{j}}\sigma
_{w_{k+1}^{i}}}+\rho _{w_{k+1}^{j},s_{k}^{i}}\dfrac{\sigma
_{s_{k}^{i}}G_{3}\left( \sigma _{w_{k+1}^{j}},\mu _{w_{k+1}^{j}}\right) }{%
\sigma _{s_{k+1}^{j}}\sigma _{s_{k+1}^{i}}\sigma _{w_{k+1}^{j}}} \\
&&+\dfrac{\mathbb{E}\left( \log \left( 1+e^{w_{k+1}^{i}}\right) \log \left(
1+e^{w_{k+1}^{j}}\right) \right) -G_{1}\left( \sigma _{w_{k+1}^{i}},\mu
_{w_{k+1}^{i}}\right) G_{1}\left( \sigma _{w_{k+1}^{j}},\mu
_{w_{k+1}^{j}}\right) }{\sigma _{s_{k+1}^{i}}\sigma _{s_{k+1}^{j}}},
\end{eqnarray*}%
where $\mathbb{E}\left( \log \left( 1+e^{w_{k+1}^{i}}\right) \log \left(
1+e^{w_{k+1}^{j}}\right) \right) $ could be computed with Gauss-Hermite
quadrature.

We have 
\begin{eqnarray*}
F\left( \sigma ,m,k\right) &\triangleq &e^{-km+\frac{k^{2}\sigma ^{2}}{2}%
}\Phi (\frac{m-k\sigma ^{2}}{\sigma }), \\
\Phi (x) &\triangleq &\dfrac{1}{\sqrt{2\pi }}\int_{-\infty }^{x}e^{-\frac{%
t^{2}}{2}}\mathrm{d}t, \\
C_{k} &\triangleq &\dfrac{\left( -1\right) ^{k+1}}{k}, \\
B_{k} &\triangleq &\frac{2\left( -1\right) ^{k+1}}{k+1}\sum_{j=1}^{k}\dfrac{1%
}{j}, \\
G_{1}\left( \sigma ,m\right) &\triangleq &m\Phi \left( \dfrac{m}{\sigma }%
\right) +\dfrac{\sigma }{\sqrt{2\pi }}e^{-\frac{m^{2}}{2\sigma ^{2}}%
}+\sum_{k=1}^{\infty }C_{k}\left[ F\left( \sigma ,m,k\right) +F\left( \sigma
,-m,k\right) \right] , \\
G_{2}\left( \sigma ,m\right) &\triangleq &\left( m^{2}+\sigma ^{2}\right)
\Phi \left( \dfrac{m}{\sigma }\right) +\left( m+\log 4\right) \dfrac{\sigma 
}{\sqrt{2\pi }}e^{-\frac{m^{2}}{2\sigma ^{2}}} \\
&&+2\sum_{k=1}^{\infty }C_{k}\left( m-k\sigma ^{2}\right) F\left( \sigma
,m,k\right) +\sum_{k=2}^{\infty }B_{k-1}\left[ F\left( \sigma ,m,k\right)
+F\left( \sigma ,-m,k\right) \right] , \\
G_{3}\left( \sigma ,m\right) &\triangleq &\sigma ^{2}\sum_{k=0}^{\infty
}\left( -1\right) ^{k}\left[ F\left( \sigma ,m,k\right) +F\left( \sigma
,m,k+1\right) \right] .
\end{eqnarray*}

\paragraph{Example 1: Exponential Trigonometric Templates, Continued}

Using the setting of Example 1, here we demonstrate an illustrative
simulation to support the fundamental of our Poisson log-normal
approximation: a sum of log-normal random vectors can be approximated by a
log-normal random vector again.

When $t=1$, a histogram of $\log \Lambda \left( B_{1},\eta \right) $ with $%
B_{1}=\left[ 0,0.1\right] $ is shown in Figure \ref{PoLNA}, where 10,000
Monte Carlo GP paths $\eta $ are drawn and we calculate%
\begin{equation*}
\Lambda \left( B_{1},\eta \right) =\int_{B_{1}}e^{\eta \left( \nu \right) }%
\mathrm{d}\nu .
\end{equation*}%
One can see that the normal distribution does quite a good job to
approximate the distribution of $\log \Lambda \left( B_{1},\eta \right) $,
which means that $\Lambda \left( B_{1},\eta \right) $ can be approximated by
a log-normal distribution. This is actually a general phenomenon, even valid
for the multivariate case.

\begin{figure}[t]
\centering\includegraphics[width=0.48\textwidth]{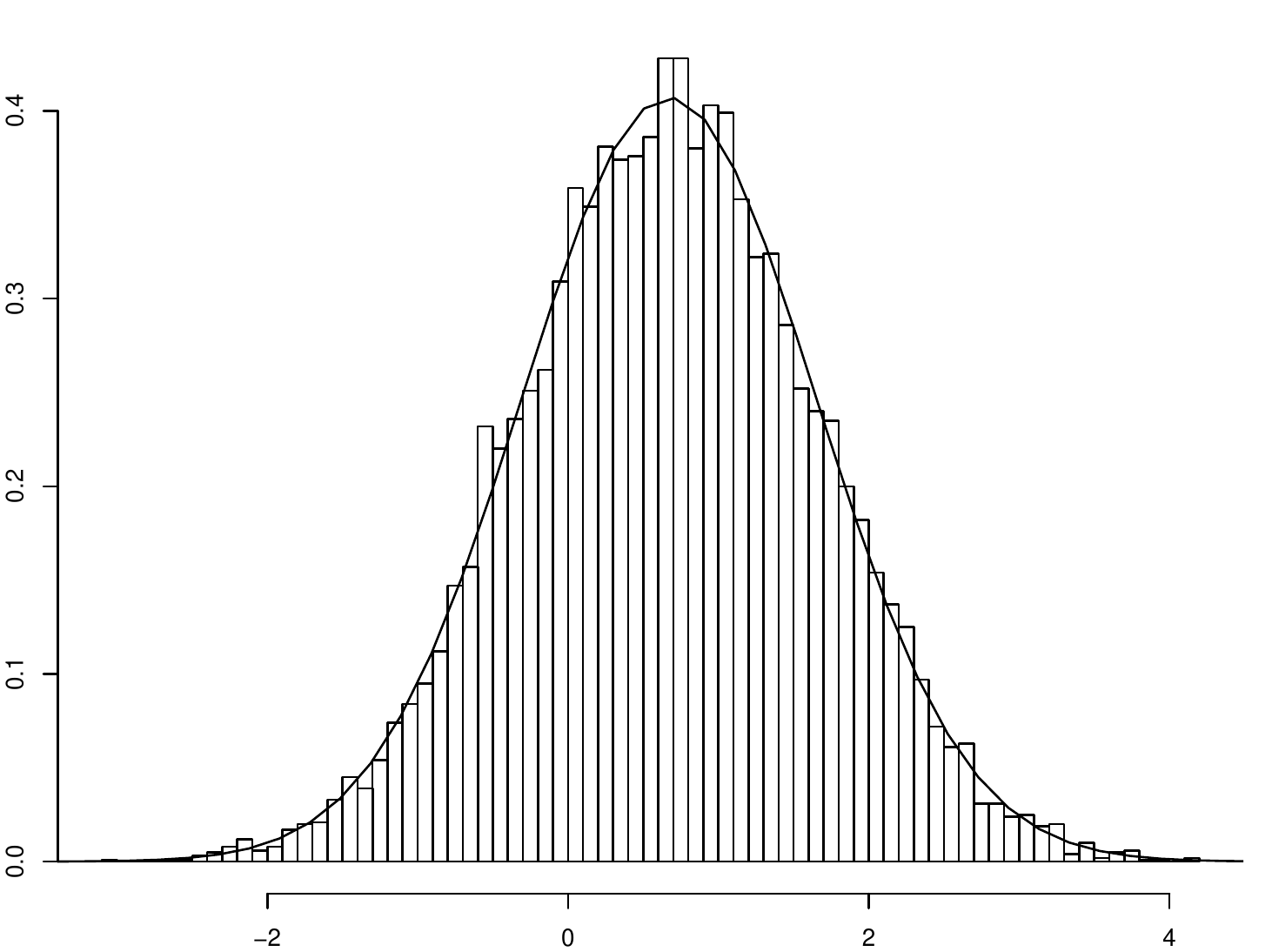}
\caption{The histogram of $\log \Lambda \left( B_{1},\protect\eta \right) $
for $B_{1}=\left[ 0,0.1\right] $ as in Example 1 with 10,000 Monte Carlo
samples. Also shown is the fitting of a normal distribution.}
\label{Fig.: PoLNA fundamental}
\end{figure}

\section{Sequential Monte Carlo Algorithm\label{Sec.:SMC alg.}}

\begin{algorithm}[h]
\caption{SMCS for BNS-ED}
\label{Main-Alg}
\begin{algorithmic}[1]
\REQUIRE

\ \

$ N$: the number of particles;

$ M_U$: the upper bound of observations;

$ M_L$: the upper bound of observations;

$ c$: the threshold of effective sample size ($\mathrm{ESS}$);

$ \epsilon $: the threshold of information gain;

$ \mathrm{Filters}$: the set of available filters and their bandwidths;

$ \mathbf{\alpha }$: the prior parameter of the model weights;

$ \tau $: the step size of the Markovian move.

\STATE Draw $\mathbf{\omega }_{0}^{\left( i\right) } \overset{\text{i.i.d.}}{\backsim} \mathrm{Dir}\left( \mathbf{\alpha }\right)$, $i=1,...,N$.

\STATE $\psi _{0}^{\left( i\right) }\leftarrow N^{-1}$, $i=1,...,N$.
\FOR{$t=1,2,3,...$}

\FOR{$i=1,...,N$, $y_{t}=M_L,...,M_U$, $\nu _{t}\in \mathrm{Filters}$}
\STATE $L_{t-1,y_{t},\nu_{t}}^{\left( i\right)}\leftarrow
        p\left( y_{t}\left\vert
                \nu _{t},\mathcal{F}_{t-1},\mathbf{\omega }_{t-1}^{\left( i\right) }
            \right.
        \right) $.
\ENDFOR

\STATE Calculate $EIG_{t}\left( \nu _{t}\right)$ with eq. (\ref{EIG_3}), $\nu _{t}\in \mathrm{Filters}$.

\STATE $\nu _{t}\leftarrow \limfunc{argmax}\limits_{\nu \in \mathrm{Filters}}EIG_{t}\left( \nu \right) $.

\STATE $y_{t}\leftarrow $ Observation at time $t $.

\STATE $\psi _{t}^{\left( i\right) }\leftarrow
        \psi _{t-1}^{\left( i\right) }\times
        L_{t-1,y_{t},\nu _{t}}^{\left( i\right)}$,
        $i=1,...,N$.

\STATE $\psi _{t}^{\left( i\right) }\leftarrow
        \psi _{t}^{\left( i\right) }/
        \sum_{i^{\prime }=1}^{N}
        \psi_{t}^{\left(i^{\prime }\right) }$,
    $i=1,...,N$.

\STATE $\mathbf{\omega }_{t}^{\left( i\right)
}\leftarrow \mathbf{\omega }_{t-1}^{\left( i\right) }$, $i=1,...,N$.

\IF{$\mathrm{ESS}\triangleq \left(
\sum_{i=1}^{N}\left( \psi _{t}^{\left( i\right) }\right) ^{2}\right) ^{-1}<c$}

\FOR{$i=1,...,N$}
\STATE Draw $\mathbf{\omega }_{t,\ast
}^{\left( i\right) }\overset{\text{i.i.d.}}{\backsim }\sum_{i^{\prime
}=1}^{N}\psi _{t}^{\left( i^{\prime }\right) }\delta _{\mathbf{\omega }_{t}^{\left( i^{\prime }\right) }}\left( \mathbf{\omega }\right) $.

\STATE $\psi _{t}^{\left( i\right)
}\leftarrow N^{-1}$.

\ENDFOR

\STATE // Markovian sampling.

\FOR{$i=1,...,N$}

\STATE Draw $\mathbf{\omega }_{t,\ast }^{\left( i\right) ,\ast }\backsim \mathrm{Dir}\left( \tau \mathbf{\omega }_{t,\ast }^{\left( i\right) }\right) $.

\STATE $A\leftarrow \mathrm{Pr}\left(
\mathbf{\omega }_{t,\ast }^{\left( i\right) }\rightarrow \mathbf{\omega }_{t,\ast }^{\left( i\right) ,\ast }\right) $.

\STATE $\mathbf{\omega }_{t}^{\left( i\right) }\leftarrow \mathbf{\omega }_{t,\ast }^{\left(
i\right) ,\ast }$ with probability $A$,
\STATE $\mathbf{\omega }_{t}^{\left( i\right) }\leftarrow \mathbf{\omega }_{t,\ast }^{\left(
i\right) }$ otherwise.

\ENDFOR

\ENDIF

\STATE $L_{t,y_{t},\nu _{t}}^{\left( i\right)
}\leftarrow p\left( y_{t}\left\vert \nu _{t},\mathcal{F}_{t-1},\mathbf{\omega }_{t}^{\left( i\right) }\right. \right) $, $i=1,...,N$.

\STATE $IG_{t}\leftarrow \sum_{i=1}^{N}\psi
_{t}^{\left( i\right) }\log \left( \dfrac{L_{t,y_{t},\nu _{t}}^{\left(
i\right) }}{\sum_{i^{\prime }=1}^{N}\psi _{t-1}^{\left( i^{\prime }\right)
}L_{t-1,y_{t},\nu _{t}}^{\left( i^{\prime }\right) }}\right) $.

\IF{$IG_{t}<\epsilon $}

\STATE Break.

\ENDIF

\ENDFOR
\end{algorithmic}
\end{algorithm}

\end{document}